\documentclass[prb,twocolumn,floatfix]{revtex4}

\usepackage{amsfonts}
\usepackage{amsmath}
\usepackage{verbatim}
\usepackage[dvips]{graphicx}
\usepackage{subfigure}

\def\Tr{\textrm}
\def\ra{\rangle}            
\def\la{\langle}            
\def\dd{\textrm{d}}
\def\Bf{\boldsymbol}

\def\ss{\scriptstyle}

\def\vxcr{b_x^{\dagger}}
\def\vycr{b_y^{\dagger}}
\def\vxan{b_x^{\phantom{\dagger}}}
\def\vyan{b_y^{\phantom{\dagger}}}

\def\rv{\Bf{r}}

\def\pv{\Bf{p}}

\begin{document}

\title{Electronic states near a quantum fluctuating point vortex in a $d$-wave superconductor:
Dirac fermion theory}
\author{Predrag Nikoli\'c, Subir Sachdev, and Lorenz Bartosch}
\affiliation{Department of Physics, Harvard University, Cambridge MA
02138, USA}

\begin{abstract}
We introduce a simple model of the low energy electronic states in
the vicinity of a vortex undergoing quantum zero-point motion in a
$d$-wave superconductor. The vortex is treated as a point flux tube,
carrying $\pi$ flux of an auxiliary U(1) gauge field, which executes
simple harmonic motion in a pinning potential. The nodal Bogoliubov
quasiparticles are represented by Dirac fermions with unit U(1)
gauge charge. The energy dependence of the local density of
electronic states (LDOS) at the vortex center has no zero bias peak;
instead, small satellite features appear, driven by transitions
between different vortex eigenmodes. These results are qualitatively
consistent with scanning tunneling microscopy measurements of the
sub-gap LDOS in cuprate superconductors. Furthermore, as argued in
L. Balents {\em et al.}, Phys.~Rev.~B {\bf 71}, 144508 (2005), the
zero-point vortex motion also leads naturally to the observed
periodic modulations in the spatial dependence of the sub-gap LDOS.
\end{abstract}

\date{July 20 2006}

\maketitle

\section{Introduction}

In a recent paper \cite{LorenzLDOS}, hereafter referred to as I, two
of the present authors considered influence of the vortex zero-point
motion on the energy dependence of local density of electronic
states (LDOS) in a $s$-wave superconductor. Here we will present a
study of vortex zero point motion in two-dimensional $d$-wave
superconductors, with an eye to application to the cuprate
supercondcutors. A direct application of the methods of paper I to
$d$-wave superconductors is discussed in Appendix~\ref{paper1}, but
this approach has significant limitations. It makes a gradient
expansion in the spatial dependence of the gap function and so
cannot be applied in the limit of small coherence length, and is not
designed to efficiently extract the important effects of the
low-energy nodal quasiparticles. The body of the present paper will
describe a new approach which overcomes these limitations.

Our focus on the zero-point motion of vortices in the cuprates is
motivated by a previous proposal that periodic modulations in the
{\em spatial dependence\/} of the LDOS inevitably appear over the
region the vortex executes its quantum zero-point motion
\cite{CompOrd1,VorMass}. Such periodic LDOS modulations have been
observed in scanning tunneling microscopy (STM) studies of the
vortex in the cuprate superconductors \cite{hoffman,fischer}. Here
we study whether the vortex motion could also explain the {\em
energy dependence\/} of the LDOS near the vortex core.

A solution of the Bogoliubov-de Gennes (BdG) equations for a vortex
in a $d$-wave superconductor leads to a large peak, as a function of
energy, at zero bias at the vortex center
\cite{Wang95,machida,Franz98}. No such peak is observed in the
experiments. The inclusion of additional density wave orders
\cite{ting,ghosal} can suppress the zero bias peak in favor of
spectral weight at satellites, but only under conditions in which
the strength of the order parameter is unacceptably large. The
theories require nearly complete magnetic/charge ordering, and there
is no indication from {\em e.g.\/} neutron scattering experiments
that such a large ordering can be present in the optimally doped
Bi-based cuprates.

Paper I extended the BdG equations to include vortex zero-point
motion for $s$-wave superconductors: it found that zero bias peak
was reduced (but not eliminated), with a transfer of spectral weight
to energies of order the vortex oscillation frequency. This was done
in an approach that performed a gradient expansion in the gap
function, which is only valid for a large core size.

The present paper examines an alternative approach to computing the
sub-gap energy dependence of the LDOS near the vortex. Our starting
point is an observation by Tsuchiura {\em et al.} \cite{ogata} that
the zero bias peak disappears when the vortex core size becomes of
order the inverse Fermi wavevector. Consequently, we will examine a
continuum model in which the vortex core becomes point-like, with a
vanishing radius. In this situation, no gradient expansion can be
performed, and it is essential to compute the effects of vortex
zero-point motion without expanding in powers of the vortex
position. We will demonstrate, within the context of a simple model,
that not only is the zero-bias peak eliminated, but the vortex
motion leads to satellite features associated with transitions
between different vortex vibrational states. We will argue that
these features are appealing candidates for explaining the sub-gap
peaks observed in STM experiments
\cite{aprile,pan,hoogen,hoffman,fischer}.

In a sense, the analysis of this paper can be viewed as a method of
computing the influence of ``phase fluctuations''\cite{phase} on the
LDOS. However, instead of explicitly integrating over the phase
degrees of freedom, we represent them by a collective co-ordinate,
the position of the vortex. Apart from the benefit of the compact
representation, this allows us to physically interpret the coupling
of the phase fluctuations to the electronic quasiparticles.
Furthermore, inclusion of Aharanov-Bohm phase factors exposes the
subtle relationship of the vortex fluctuations to competing order
parameters \cite{CompOrd1}.

The outline of the paper is as follows. Our model will be introduced
in Section~\ref{sec:model}, along with an initial discussion of its
characteristic properties. The details of our calculations are
presented in Section~\ref{sec:pert}. After setting up the formalism,
we first analyze the LDOS influenced only by the vortex zero-point
quantum motion, and then include the resonant scattering at the
lowest order of perturbation theory. Plots of the full LDOS are
shown in the Sections~\ref{sec:plots} and~\ref{sec:magnus}. We
summarize our results and their relation to experiments in
Section~\ref{sec:conc}.

\section{The model}
\label{sec:model}

We consider a single vortex coupled to quasiparticles in a clean
$d$-wave superconductor at zero temperature. The vortex is assumed
to experience a harmonic trapping potential in which it can carry
out its quantum zero-point motion. Such a potential can result from
interactions between vortices in a vortex lattice, or from a pinning
impurity. The Hamiltonian can be written as ($\hbar=1$):
\begin{eqnarray}
H &=& \frac{\pv_\Tr{v}^2}{2m_\Tr{v}} + \frac{1}{2} m_\Tr{v}
\omega_\Tr{v}^2 \rv_\Tr{v}^2 \nonumber \\
&~&~~~~+ \sum_{\Tr{nodes}}
    \int \dd^2 r \Psi^{\dagger}(\rv)  H_{BdG}(\rv) \Psi^{\phantom{\dagger}}(\rv) \ .
    \label{Hamiltonian}
\end{eqnarray}
For now we neglect the Magnus force on the vortex, assuming that it
is much smaller that the trapping force; we shall consider it
carefully in the Section~\ref{sec:magnus}. The operators
$\pv_\Tr{v}$ and $\rv_\Tr{v}$ are the canonical momentum and
position of the vortex. For simplicity, we place the vortex in an
isotropic trap characterized by a single harmonic frequency
$\omega_\Tr{v}$. It has been shown that presence of static vortices
does not qualitatively change the low-energy spectrum of nodal
$d$-wave quasiparticles \cite{Franz98,Tesh1,Tesh2}, and in this
paper we will find that the gapless nodes also survive quantum
fluctuations of vortices. Therefore, quasiparticles are massless
Dirac spinors and we describe them by Nambu operators
$\Psi^{\dagger}(\rv)$ and $\Psi(\rv)$ defined at every node in the
$d$-wave spectrum. Being interested only in the low-energy dynamics,
we linearize the Bogoliubov-de Gennes Hamiltonian in the vicinity of
gap-nodes, and apply a convenient Franz-Te\v sanovi\'c unitary
transformation. For the node $\Bf{p} \approx k_{\textrm{f}}
\hat{\Bf{x}}$ we have \cite{Tesh1,Tesh2}:
\begin{equation}\label{BdG}
H_{\textrm{BdG}} = v_{\textrm{f}} (p_x + a_x) \sigma^z +
      v_{\Delta} (p_y + a_y) \sigma^x + m v_{\textrm{f}} v_x I \ ,
\end{equation}
and we need not explicitly worry about the other nodes because their
linearized Hamiltonians are related by unitary transformations. Note
that we formally work with $d_{xy}$ symmetry, which is related to
$d_{x^2-y^2}$ by a rotation. In this expression, $\Bf{p}$ is the
quasiparticle momentum operator relative to the node, $m$ is the
electron mass, $v_{\textrm{f}}$ and $v_{\Delta}$ are the Fermi and
gap velocities respectively, and $\sigma^\mu$ are Pauli matrices.
The effective gauge field $\Bf{a}$ is proportional to the phase
gradient of the superconducting order parameter, and thus
corresponds to the $\pi$-flux centered at the vortex location. The
role of $\Bf{a}$ is to implement the statistical interaction between
vortices and quasiparticles, so that when a quasiparticle completes
a circle around the vortex its wavefunction changes sign. The
supercurrent velocity field $\Bf{v}$ appears in the Hamiltonian
because the supercurrents that circulate around the vortex give rise
to Doppler shifts of quasiparticle energies. Note, however, that
only the projection of $\Bf{v}$ on the nodal direction matters, and
also that Doppler shift effects decrease rapidly beyond the London
penetration depth from the vortex center.

Our goal is to elucidate only the qualitative features of
quasiparticle spectra in the vicinity of a fluctuating vortex.
Hence, we will make several simplifications that are not
quantitatively justified in realistic situations, but the obtained
spectra will nevertheless be remarkably similar to the spectra
observed in experiments. The complexity of calculations is
considerably reduced if the Doppler shift and all effects of
anisotropy are neglected. The Bogoliubov-de Gennes Hamiltonian near
the node $\Bf{p} \approx \pm k_{\textrm{f}} \hat{\Bf{x}}$ becomes:
\begin{equation}\label{IsoBdG}
H_{BdG} = \left(
  \begin{array}{cc}
    p_x+a_x & p_y+a_y  \\
    p_y+a_y & -(p_x+a_x)
  \end{array}
\right) \ ,
\end{equation}
where we have set $v_{\textrm{f}}$ and $v_{\Delta}$ to unity. We
will work in the units $\hbar = v_{\textrm{f}} =1$ at all times,
except when we discuss scales.

The coupling between the fermions and the vortex arises from the
gauge field. This is specified by the instantaneous position
operator $\rv_\Tr{v}$ of the vortex:
\begin{equation}\label{VGF}
\Bf{a}(\rv,\rv_\Tr{v}) =
  \frac{\hat{\Bf{z}} \times (\rv - \rv_\Tr{v})}{2 |\rv - \rv_\Tr{v}|^2} \ ,
\end{equation}
assuming that we can regard the vortex core to be negligibly small.

The physical reason for these simplifications comes from our
expectation that it is the vortex quantum fluctuations that produce
the sub-gap peaks at about 7 meV from the Fermi level in the
quasiparticle density of states \cite{aprile,pan,hoogen}. Even if we
regard the core of a static vortex to be infinitely small, as was
done in the expression ~(\ref{VGF}), the zero-point quantum motion
of the vortex in the harmonic trap creates a finite central region
in which the supercurrents are gradually suppressed. The circulating
superflow is then the strongest at some finite radius.
Quasiparticles can scatter from this supercurrent distribution
through any number of virtual excited states of the oscillating
vortex. The result is a weak resonant state, a metastable binding
between the vortex and a quasiparticle. In our simplified
calculation this effect is solely due to the electron Berry phase,
but the Doppler shift should also contribute it, even more directly
as a partially attractive potential. Indeed, our calculation will
qualitatively reproduce the sub-gap peaks in the density of states,
which are reminiscent of a localized metastable state in the vortex
core, and in our model originate from resonant scattering. In the
light of this scenario, we do not expect anisotropies to play a
qualitatively significant role. The Doppler shift may be important
in enhancing these effects, but the mere demonstration of the
sub-gap peaks through the Berry phase alone makes the case for
relevance of the vortex quantum motion. Another possible mechanism
is resonant scattering from the conventional vortex core, where the
superconducting order parameter magnitude is suppressed. However, in
$d$-wave superconductors the core is very small (a few lattice
spacings), so that the corresponding resonance would occur only at
relatively large energy. Note that we are not considering the
possibility of bound core states: all quasiparticle states are
expected to be extended in pure $d$-wave superconductors, even in
the presence of vortices \cite{Franz98,Tesh1,Tesh2}.

Our main result is the quasiparticle local density of states (LDOS)
in the vicinity of a quantum oscillating vortex. We calculate the
LDOS perturbatively, using
\begin{equation}\label{SmallParameter}
\alpha = \left( \frac{m_\Tr{v} v_\Tr{f}^2}{\hbar \omega_\Tr{v}}
\right)^\frac{1}{2}
\end{equation}
as a small dimensionless parameter. We treat the vortex effective
mass $m_\Tr{v}$ as an independent quantity, although it can be
determined microscopically \cite{DWVorAct}. After all
simplifications, we can write the LDOS as the following scaling
function of energy $\epsilon$ and distance $r$ from the origin
(center of the vortex trap):
\begin{equation}\label{LDOSscaling}
\rho(\epsilon,r) = \frac{\omega_\Tr{v}}{\hbar v_\Tr{f}^2}
\sum_{n=0}^{\infty}
  \alpha^{2n} F_n \left( \frac{\epsilon}{\hbar \omega_\Tr{v}},
             \frac{\epsilon r}{\hbar v_\Tr{f}} ; \alpha \right) \ .
\end{equation}
The universal functions $F_n(x,y;\alpha)$ can be regarded as being
of the same order of magnitude for any given finite arguments. Note,
however, that this is not a Taylor expansion. The extra dependence
of $F_n$ on $\alpha$ is a somewhat unusual consequence of the proper
choice of perturbation. It is usual to start from a non-interacting
theory and include all interactions as perturbations. Unfortunately,
interactions between quasiparticles and vortices are not weak, and
cannot be characterized by a small expansion parameter. Instead, it
is better to include the influence of the vortex zero-point quantum
motion on quasiparticles at the zeroth order, and treat the resonant
scattering of quasiparticles from the oscillating vortex as a
perturbation. Thus we perform a partial resummation of the $\alpha$
expansion, and this accounts for the $\alpha$ dependence of the
co-efficients in Eq.~(\ref{LDOSscaling}); features of the zero-point
vortex motion, also parameterized by $\alpha$, appear in the LDOS at
all orders of perturbation theory. The small parameter $\alpha$ is
the ratio of such perturbation's energy scale and the energy barrier
$\hbar \omega_\Tr{v}$ for virtual transitions.

As a matter of fact, only the zero-point vortex oscillations set the
scale for the zeroth order term in ~(\ref{LDOSscaling}), which
results in an additional scaling property of $F_0(x,y;\alpha)$:
\begin{equation}\label{LDOS0scaling}
F_0(x,y;\alpha) = \alpha F_0(x/\alpha,y;1) \ .
\end{equation}
One should not be misled to conclude that $F_0$ becomes small when
$\alpha \to 0$: as we will show, $F_0(x,y;\alpha) \to x/(2\pi)$ for
$x \gg \alpha$.

\section{Perturbation theory}
\label{sec:pert}

The Hamiltonian ~(\ref{Hamiltonian}) can be expressed in a fashion
analogous to the Holstein-Primakoff expansion for the spin systems.
In absence of quasiparticles the vortex is modeled by a
two-dimensional linear harmonic oscillator, whose eigenstates
$|n_x,n_y\ra$ are characterized by two integer quantum numbers,
$n_x$ and $n_y$. Let us define the following ``matrix elements'':
\begin{equation}\label{HMatrEl}
V_{(n_{1x},n_{1y}),(n_{2x},n_{2y})}(\rv) = \la n_{1x},n_{1y} |
  H_{BdG}(\rv) | n_{2x},n_{2y} \ra \ .
\end{equation}
We also introduce creation operators, $\vxcr$ and $\vycr$, which
raise the vortex quantum number:
\begin{equation}
\vxcr |n_x,n_y \ra = \sqrt{n_x + 1} |n_x + 1, n_y \ra,
\end{equation}
and similarly for $\vycr$. If we insert the identity operators on
the left and right side of $H_{BdG}$ in the equation
~(\ref{Hamiltonian}), and resolve them in terms of the vortex
eigenstates $|n_x,n_y\ra$, we can systematically write:
\begin{eqnarray}\label{PertExp}
&& \Psi^{\dagger} H_{BdG} \Psi^{\phantom{\dagger}} = \Psi^{\dagger}
\Bigl\lbrace
          V_{(0,0),(0,0)} + \\
&& ~~~~~ V_{(1,0),(0,0)} \vxcr + h.c. + V_{(0,1),(0,0)} \vycr + h.c. + \nonumber \\
&& ~~~~~ \left( V_{(1,0),(1,0)} - V_{(0,0),(0,0)} \right) \vxcr\vxan + \nonumber \\
&& ~~~~~ \left( V_{(0,1),(0,1)} - V_{(0,0),(0,0)} \right) \vycr\vyan + \nonumber \\
&& ~~~~~ V_{(1,0),(0,1)} \vxcr\vyan + h.c. +
          V_{(1,1),(0,0)} \vxcr\vycr + h.c. + \nonumber \\
&& ~~~~~ \frac{1}{\sqrt{2}} V_{(2,0),(0,0)} \vxcr\vxcr + h.c. + \nonumber \\
&& ~~~~~ \frac{1}{\sqrt{2}} V_{(0,2),(0,0)} \vycr\vycr + h.c. +
         \cdots \Bigr\rbrace  \Psi^{\phantom{\dagger}} \nonumber \ .
\end{eqnarray}
It can be easily seen that the quasiparticle momentum operators
$\Bf{p}$ appear only in the lowest order term $V_{(0,0),(0,0)}$. In
all elastic processes (such as $\vxcr\vxan$) the momentum operators
that appear in the matrix elements $V$ are cancelled out. On the
other hand, in all inelastic processes the momentum operators do not
appear because they do not excite vortex states. Therefore, for
practical purposes we can replace all matrix elements except
$V_{(0,0),(0,0)}$ with:
\begin{equation}\label{HMatrEl2}
V_{(n_{1x},n_{1y}),(n_{2x},n_{2y})}(\rv) =
  \left(
    \begin{array}{cc}
      a_x'(\rv) & a_y'(\rv)  \\
      a_y'(\rv) & -a_x'(\rv)
    \end{array}
  \right) \ ,
\end{equation}
where
\begin{eqnarray}\label{EffVGF}
\Bf{a}'(\rv) & \equiv & \Bf{a}_{(n_{1x},n_{1y}),(n_{2x},n_{2y})}(\rv) \\
& = & \la n_{1x},n_{1y} | \Bf{a}(\rv,\rv_\Tr{v}) | n_{2x},n_{2y} \ra
\ . \nonumber
\end{eqnarray}
Note that the bare effective gauge field $\Bf{a}(\rv,\rv_\Tr{v})$,
given by ~(\ref{VGF}), does not define any energy or length scale.
The scales are introduced through the states $|n_x,n_y\ra$ of the
quantum harmonic oscillator. The harmonic oscillator wavefunctions
depend only on the dimensionless coordinates $\Bf{\xi} =
\sqrt{m_\Tr{v} \omega_\Tr{v}} \Bf{r}$, so that the energy scale
associated with $V_{(n_{1x},n_{1y}),(n_{2x},n_{2y})}$ is $(\hbar
\omega_\Tr{v} m_\Tr{v} v_\Tr{f}^2)^{1/2}$ (when all physical
constants are put in their places).

The expansion ~(\ref{PertExp}) separates various processes in which
quasiparticles can scatter elastically or inelastically from a
vortex. At low enough energies the expansion can be truncated after
a few lowest order terms. In the following we will set up a
perturbation theory, with the unperturbed Hamiltonian:
\begin{equation}\label{H0}
H_0 =  \omega_\Tr{v} \left( \vxcr\vxan + \vycr\vyan \right) +
       \int \dd^2 r \Psi^{\dagger}  V_{(0,0),(0,0)}  \Psi^{\phantom{\dagger}} \ ,
\end{equation}
and the perturbation:
\begin{equation}\label{H1}
H_1 = \int \dd^2 r \Psi^{\dagger}
     \left( V_{(1,0),(0,0)} \vxcr + V_{(0,1),(0,0)} \vycr + h.c. \right)
     \Psi^{\phantom{\dagger}} \ .
\end{equation}
The more complicated scattering processes in ~(\ref{PertExp}) become
qualitatively important only at energies $\epsilon \sim
2\omega_\Tr{v}$ and above.

Before we begin calculations, it is useful to identify the small
parameter of the perturbation theory. The available parameters in
the model ~(\ref{Hamiltonian}) define two energy scales: the energy
of vortex quantum oscillations $\hbar \omega_\Tr{v}$, and the
characteristic kinetic energy $m_\Tr{v} v_\Tr{f}^2$. The unperturbed
Hamiltonian ~(\ref{H0}) is properly characterized by the scale
$|H_0| = \hbar \omega_\Tr{v}$, because the vortex harmonic frequency
sets the energy barrier that needs to be crossed in perturbative
virtual transitions. The energy scale of the perturbation
~(\ref{H1}) is $|H_1| = (\hbar \omega_\Tr{v} m_\Tr{v}
v_\Tr{f}^2)^{1/2}$, as discussed above. Therefore, we can define the
small parameter $\alpha = |H_1|/|H_0|$ given by expression
~(\ref{SmallParameter}). The lowest order correction $\rho_1$ to the
quasiparticle LDOS comes from the Fock exchange process
(self-energy), so that $\rho_1 \propto \alpha^2$. Namely, every
vertex $\propto |H_1|$ in Feynman diagrams introduces a factor of
$\alpha$, which becomes apparent after rescaling all energies by
$|H_0| = \hbar \omega_\Tr{v}$ as was done in ~(\ref{LDOSscaling}).

In practice, one also has to keep in mind the quasiparticle cut-off
energy $\Lambda$. This third energy scale physically comes from the
superconducting gap amplitude, beyond which linearization of the
Bogoliubov-de Gennes Hamiltonian breaks down. Furthermore,
renormalization of the effective vortex mass due to quasiparticles
\cite{DWVorAct} is such that $m_\Tr{v} v_\Tr{f}^2 \sim \Lambda$.
Irrespective of its physical origin, numerical analysis requires
that $\Lambda$ be introduced. However, in the ideal linearized model
that we are concerned with, we can in principle take the $\Lambda
\to \infty$ limit. As long as we treat the vortex mass $m_\Tr{v}$
and frequency $\omega_\Tr{v}$ as independent parameters, we find
that the spectrum does not depend on the numerically introduced
$\Lambda$ at energies sufficiently below $\Lambda$. The functions
$F_n$ in ~(\ref{LDOSscaling}) are universal.

\subsection{The LDOS due to vortex zero-point quantum motion}
\label{sec:ldos0}

We begin by considering the lowest order term in the expansion
~(\ref{LDOSscaling}) for the LDOS:
\begin{equation}\label{Rho0}
\rho_0(\epsilon,r) = \frac{\omega_\Tr{v}}{\hbar v_\Tr{f}^2}
   F_0 \left( \frac{\epsilon}{\hbar \omega_\Tr{v}},
             \frac{\epsilon r}{\hbar v_\Tr{f}} ; \alpha \right) \ .
\end{equation}
This term contains effects of the vortex zero-point quantum motion
on quasiparticle spectra, but excludes effects of the resonant
scattering. For this purpose we will numerically diagonalize the
Hamiltonian ~(\ref{H0}) in infinite space.

It will be useful to first understand the solution for a static
vortex, which is obtained in the limit of $m_\Tr{v} \omega_\Tr{v}
\to \infty$. In this limit there are no vortex fluctuations, so that
$V_{(0,0),(0,0)}$ reduces to ~(\ref{IsoBdG}), with the gauge field
given by ~(\ref{VGF}), and $\rv_\Tr{v} \equiv 0$. Exact analytical
solutions (properly normalized in the infinite space) are then found
to be \cite{DWVorAct}:
\begin{eqnarray}\label{WFinf}
&& \psi_{q,l,k}^\infty(r,\phi) = \sqrt{\frac{\epsilon}{4\pi}} \times \\
&& ~~ \left\lbrace
  \begin{array}{lcl}
    \left(
      \begin{array}{c}
        J_{-l+\frac{1}{2}}(kr) e^{i(l-1)\phi} \\
        -iq \cdot J_{-l-\frac{1}{2}}(kr) e^{il\phi}
      \end{array}
    \right)
    & \quad , \quad & l \leq 0 \\
    \left(
      \begin{array}{c}
        J_{l-\frac{1}{2}}(kr) e^{i(l-1)\phi} \\
        iq \cdot J_{l+\frac{1}{2}}(kr) e^{il\phi}
      \end{array}
    \right)
    & \quad , \quad & l \geq 0
  \end{array}
\right\rbrace \nonumber
\end{eqnarray}
in the representation that diagonalizes the angular momentum.
Quasiparticles carry the following quantum numbers: ``charge'' $q =
\pm 1$ (distinguishes particle-like and hole-like states), angular
momentum $l \in \mathbb{Z}$ and radial wavevector $k>0$. There are
only extended quasiparticle states with energies $\epsilon = q |k|$.
$J_l(kr)$ are Bessel functions of the first kind. The wavefunction
for $l=0$ is a linear combination of the two forms of
$\psi_{q,0,k}^\infty$ in ~(\ref{WFinf}), which can be uniquely
specified only with a separate boundary condition at the vortex
location (an artifact of $m_\Tr{v} \omega_\Tr{v} \to \infty$ and a
point-like core) \cite{Tesh2}. The corresponding LDOS $\rho_\infty$
diverges as $1/r$ at any energy, due to the $l=0$ wavefunctions,
which include $J_{-\frac{1}{2}}(kr)$:
\begin{eqnarray}\label{LDOSstatic}
\rho_\infty (\epsilon, r) & = &
   \frac{\cos(2|\epsilon|r)}{2 \pi^2 r} + \frac{|\epsilon|}{\pi} \sum_{l=0}^{\infty}
   J_{l+\frac{1}{2}}^2 (|\epsilon|r) \nonumber \\
& \longrightarrow & \left\lbrace
       \begin{array}{c@{\quad , \quad}c}
          \frac{1}{2 \pi^2 r} & |\epsilon|r \ll 1 \\[2mm]
          \frac{|\epsilon|}{2\pi} & |\epsilon|r \gg 1
       \end{array} \right\rbrace  \ .
\end{eqnarray}

For finite $\omega_\Tr{v}$ and $m_\Tr{v}$ the vortex fluctuates in a
finite region whose radius is $\sim (\omega_\Tr{v}
m_\Tr{v})^{-1/2}$. This is a property of the harmonic oscillator
ground-state. The exact quasiparticle eigenstates are still
characterized by the same quantum numbers, and only the dependence
on radius $r$ is changed. This is so because in the ground-state of
our model the vortex wavefunction is rotationally symmetric, and
well localized in a finite region of space. The LDOS $\rho_0$
affected only by vortex zero-point oscillations will coincide with
~(\ref{LDOSstatic}) at energies $\epsilon \gg (m_\Tr{v}
\omega_\Tr{v})^{1/2}$ or distances $r \gg (m_\Tr{v}
\omega_\Tr{v})^{-1/2}$.

The first step in diagonalizing ~(\ref{H0}) is to calculate
$V_{(0,0),(0,0)}$. From the definition ~(\ref{HMatrEl}) we see that
$V_{(0,0),(0,0)}$ is given by the expression ~(\ref{IsoBdG}) with an
effective ``gauge field'':
\begin{equation}
\Bf{a}_{(0,0),(0,0)}(\rv) = \la 0,0 | \Bf{a}(\rv,\rv_\Tr{v}) | 0,0
\ra \ ,
\end{equation}
where $\Bf{a}(\rv,\rv_\Tr{v})$ is given by ~(\ref{VGF}).
Substituting here the ground-state wavefunctions of the linear
harmonic oscillator, we get:
\begin{equation}
\Bf{a}_{(0,0),(0,0)}(\rv) = \left( 1 - e^{-m_\Tr{v} \omega_\Tr{v}
r^2} \right) \Bf{a}(\rv,0) \ .
\end{equation}
Details of this derivation are presented in Appendix
~\ref{appVertex}. The effective gauge field is not singular at the
origin, even though the static vortex has a point-like core. This
indicates that supercurrents are suppressed near the origin.
Effectively, a ``vortex core'' has been created, and hence the exact
wavefunctions are significantly modified from ~(\ref{WFinf}).
Behavior of the exact wavefunctions $\psi = ( e^{i(l-1)\phi} u(r) ,
e^{i l \phi} v(r) )$  at small $r$ becomes:
\begin{equation}
u(r) \propto J_{l-1} (\kappa_l r) \ ,
\end{equation}
and similar for $v(r)$, where $\kappa_l = \sqrt{k^2 - l m_\Tr{v}
\omega_\Tr{v}}$. These wavefunctions are always finite or zero at
the origin, so that the LDOS is not divergent. For $r \gg
(\omega_\Tr{v} m_\Tr{v})^{-1/2}$ the exact wavefunctions reduce to
the form given by ~(\ref{WFinf}), except that there is a phase
shift: $\psi(r,\phi) = \psi^\infty(r+\delta_{l,k},\phi)$. The known
behavior for small and large $r$ allows one to easily find and
normalize exact wavefunctions in infinite space: the radial
Schr\"odinger equation is first numerically solved in a finite
region of space $r < r_0$, and then the solution is matched by value
and derivative to the phase shifted form ~(\ref{WFinf}) for $r > r_0
\gg (m_\Tr{v} \omega_\Tr{v})^{-1/2}$.

\begin{figure}
\includegraphics[width=3.2in]{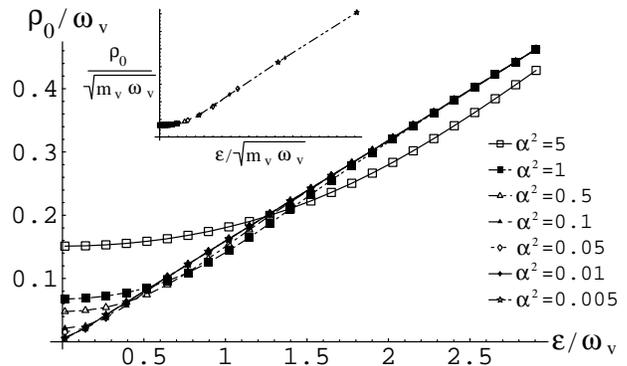}
\caption{\label{LDOS0}The quasiparticle LDOS influenced solely by
the vortex zero-point quantum motion. The LDOS is plotted at the
origin as a function of energy and $\alpha^2$. All plots can be
rescaled to a single curve, shown in the inset.}
\end{figure}

\begin{figure}
\includegraphics[width=3.0in]{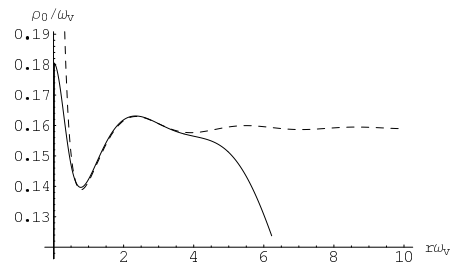}
\caption{\label{LDOS0vsR}The quasiparticle LDOS influenced solely by
the vortex zero-point quantum motion, plotted at the energy
$\epsilon = \omega_\Tr{v}$ as a function of distance $r$ from the
origin. The solid and dashed lines correspond to an oscillating
($\alpha^2=5$), and a static vortex respectively. The apparent drop
of LDOS for the oscillating vortex beyond $r \approx 4$ is an
artifact of numerical calculation, which took into account only a
small number of angular momentum channels ($-5 \leq l \leq 5$).}
\end{figure}

In the figure ~\ref{LDOS0} we plot the LDOS $\rho_0(\epsilon)$ at
the origin for several values of the dimensionless parameter
$\alpha^2$. All plots collapse to a single curve (shown in the
inset) if both the energy and LDOS are rescaled by $(m_\Tr{v}
\omega_\Tr{v})^{1/2}$, because only this energy appears in the
quasiparticle part of the Hamiltonian ~(\ref{H0}). Hence, the
following scaling applies:
\begin{equation}
\rho_0(\epsilon,r) =
   \frac{\sqrt{\hbar m_\Tr{v} \omega_\Tr{v} v_\Tr{f}^2}}{(\hbar v_\Tr{f})^2}
   F_0' \left( \frac{\epsilon}{\sqrt{\hbar m_\Tr{v} \omega_\Tr{v} v_\Tr{f}^2}},
             \frac{\epsilon r}{\hbar v_\Tr{f}} \right) \ ,
\end{equation}
which in comparison with ~(\ref{Rho0}) implies:
\begin{equation}
F_0(x,y;\alpha) = \alpha F_0'(x/\alpha,y) = \alpha F_0(x/\alpha,y;1)
\ .
\end{equation}
At large energies ($\epsilon \gg (m_\Tr{v} \omega_\Tr{v})^{1/2}$),
the LDOS at the origin begins to reflect the bulk density of states,
and hence becomes a linear function of energy. Furthermore,
$\rho_0(\epsilon,r)$ in the bulk ($r \gg (m_\Tr{v}
\omega_\Tr{v})^{-1/2}$) must coincide with $\rho_\infty(\epsilon,
r)$ of the static vortex, so that the expression ~(\ref{LDOSstatic})
determines both the large energy and large distance behavior of
$\rho_0(\epsilon,r)$:
\begin{equation}
F_0(x,y;\alpha) \to \frac{x}{2\pi} \quad , \quad \frac{x}{\alpha}
\vee y \gg 1 \ .
\end{equation}
Spatial variations of the LDOS are illustrated in the figure
~\ref{LDOS0vsR}. Zero-point quantum motion of the vortex removes the
$1/r$ divergence of the LDOS found in the core of a static vortex,
while at large distances quasiparticles do not feel the vortex
fluctuations.

Smallness of the vortex core, measured by the coherence length $\xi
\ll (m_\Tr{v} \omega_\Tr{v})^{-1/2}$, seems to be responsible for
absence of the zero-energy peak in the quasiparticle LDOS, as was
suggested in the Ref.\onlinecite{ogata}. This can be inferred by
comparing the LDOS obtained here with the LDOS of a similar calculation from the Appendix~\ref{paper1}. The approach from the Appendix works in the opposite limit $\xi > (m_\Tr{v} \omega_\Tr{v})^{-1/2}$ and produces a broad zero-energy peak in the quasiparticle LDOS.

\subsection{One-loop correction}
\label{sec:oneloop}

In order to explore the resonant scattering of quasiparticles from
the fluctuating vortex, we calculate the one-loop correction to the
quasiparticle Green's function. This will directly lead to the
lowest order LDOS correction:
\begin{equation}
\rho_1(\epsilon,r) = \alpha^2 \frac{\omega_\Tr{v}}{\hbar v_\Tr{f}^2}
   F_1 \left( \frac{\epsilon}{\hbar \omega_\Tr{v}},
              \frac{\epsilon r}{\hbar v_\Tr{f}} ; \alpha \right) \ ,
\end{equation}
which we will discuss in Section~\ref{sec:plots}

The Hamiltonian given by ~({\ref{H0}) and ~(\ref{H1}) defines a
problem of massless Dirac fermions coupled to a single bosonic
oscillator. The unperturbed fermion states are characterized by
``charge'' $q$, radial wavevector $k$ and angular momentum $l$.
Perturbation theory can be visualized using the standard Feynman
diagram technique. The bare propagators ($G_0$ for quasiparticles
and $D_0^{\mu\nu}$ for the vortex) and vertices are:
\begin{equation}\label{PropVert}
    \begin{array}{ccc}
        \includegraphics[width=0.8in]{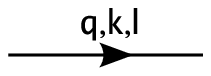}  & \qquad &
              G_0 (q,k,l,\omega) = \frac{1}{\omega - q (k - i 0^+)} \\
        \includegraphics[width=0.8in]{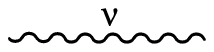}  & \qquad &
              D_0^{\mu\nu} (\omega) = \frac{2\omega_\Tr{v}}{\omega^2 - \omega_\Tr{v}^2
              + i 0^+} \delta_{\mu\nu} \\
        \begin{minipage}{1in}
            \includegraphics[width=1in]{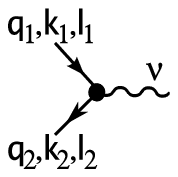}
        \end{minipage}
             & \qquad &
              V_{q_1,l_1,k_1;q_2,l_2,k_2}^{\nu}
    \end{array}
\end{equation}
Here, the indices $\mu$ and $\nu$ denote spatial directions $x$ and
$y$, and summation over repeated indices will be assumed from now
on. The vertex operator $V_{q_1,l_1,k_1;q_2,l_2,k_2}^{\mu}$ is the
Fourier transform of $V_{(1,0),(0,0)} = V_{(0,0),(1,0)}$ for $\mu=x$
and $V_{(0,1),(0,0)} = V_{(0,0),(0,1)}$ for $\mu=y$. Frequency is
conserved at vertices, but $k$ and $l$ are not. Physical
conservation of angular momentum is reflected by requirement that
$V_{q_1,l_1,k_1;q_2,l_2,k_2}^{\mu} \propto \delta_{|l_1-l_2|,1}$;
the first excited states of the 2D harmonic oscillator carry angular
momentum $l = \pm 1$, and in these inelastic scattering processes
the quasiparticle angular momentum changes by one.

The one-loop self-energy $\Sigma$ has two contributions, in
principle:
\begin{equation}
    \begin{array}{ccc}
        \includegraphics[width=1in]{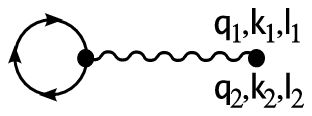} & \qquad &
        \includegraphics[width=1.5in]{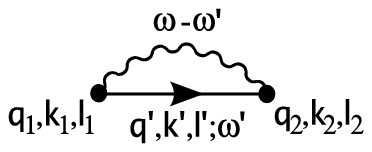} \\
        \textrm{Hartree} & \qquad & \textrm{Fock}
    \end{array}
\end{equation}
The Hartree self-energy is zero within the present approximations
because the vertex that touches the fermion loop has the same
incoming and outgoing angular momentum. The Fock self-energy is:
\begin{eqnarray}\label{SEfock}
& & \Sigma_{q_1,l_1,k_1;q_2,l_2,k_2}(\omega) =
      i \sum_{q',l'} \int \dd k' \frac{\dd\omega'}{2\pi} G_0(q',l',k',\omega')
      \nonumber \\
& & ~~~ \times D_0^{\mu\nu}(\omega-\omega')
         V_{q_1,l_1,k_1;q',l',k'}^\mu V_{q',l',k';q_2,l_2,k_2}^\nu \nonumber \\
& & ~~~ = \sum_{q',l'} \int \dd k'
        \frac {V_{q_1,l_1,k_1;q',l',k'}^\mu V_{q',l',k';q_2,l_2,k_2}^\mu}{\omega -
        q'(\omega_\Tr{v} + k' - i0^+)} \ .
\end{eqnarray}
The one-loop correction to the quasiparticle Green's function is
given by the simplest diagram that contains only one self-energy
part:
\begin{eqnarray}\label{G1}
&& G_{q_1,l_1,k_1;q_2,l_2,k_2}^{(1)}(\omega) = \\
&& ~~~~~ G_0(q_1,l_1,k_1,\omega) G_0(q_2,l_2,k_2,\omega)
    \Sigma_{q_1,l_1,k_1;q_2,l_2,k_2}(\omega) \ . \nonumber
\end{eqnarray}

The remaining task is to calculate the vertex operators
$V_{q_1,l_1,k_1;q_2,l_2,k_2}^{\mu}$ in momentum space. In the
original position representation these operators have the form
~(\ref{HMatrEl2}), where the first-order effective gauge fields
$\Bf{a}^\mu$ ($\mu \in \lbrace x,y \rbrace$) are:
\begin{eqnarray}\label{VGF1}
\Bf{a}^x & = & \la 1,0 | \Bf{a}(\rv,\rv_\Tr{v}) | 0,0 \ra =
  - \sqrt{\frac{m_\Tr{v} \omega_\Tr{v}}{2}} \times \\
& \times & \left\lbrack
      \frac{1-e^{-\xi^2}}{\xi^2} \hat{\Bf{y}}
    + \frac{(1+\xi^2)e^{-\xi^2} - 1}
           {\xi^2} \cdot \frac{2\xi_x \hat{\Bf{z}}\times\Bf{\xi}}{\xi^2}
    \right\rbrack \nonumber \\
\Bf{a}^y & = & \la 0,1 | \Bf{a}(\rv,\rv_\Tr{v}) | 0,0 \ra =
    \sqrt{\frac{m_\Tr{v} \omega_\Tr{v}}{2}} \times \nonumber \\
& \times & \left\lbrack
      \frac{1-e^{-\xi^2}}{\xi^2} \hat{\Bf{x}}
    - \frac{(1+\xi^2)e^{-\xi^2} - 1}
           {\xi^2} \cdot \frac{2\xi_y \hat{\Bf{z}}\times\Bf{\xi}}{\xi^2}
    \right\rbrack \nonumber \ .
\end{eqnarray}
Here, $\Bf{\xi} = (\xi_x,\xi_y)$ are the usual dimensionless
coordinates of the quantum harmonic oscillator, given by $\Bf{\xi} =
\sqrt{m_\Tr{v} \omega_\Tr{v}} \Bf{r}$. For switching to the momentum
representation it will be convenient to keep the spinor structure
and organize various quantities into spinor matrices. The role of
spin $s$ is taken over by the quantum number $q$ in the momentum
representation, which can take only two values being the sign of
energy. First, we define the Fourier weight matrix, which is used to
translate between the position and momentum representations:
\begin{equation}
T_{l,k}(\rv) = \left(
  \begin{array}{cc}
    u_{+,l,k}(\rv) & u_{-,l,k}(\rv) \\
    v_{+,l,k}(\rv) & v_{-,l,k}(\rv)
  \end{array}
\right) \ .
\end{equation}
This matrix is expressed in terms of the eigenfunctions of the
unperturbed Hamiltonian ~(\ref{H0}), which in the position
representation look like:
\begin{equation}\label{WF}
\psi_{q,k,l}(\rv) = \left(
  \begin{array}{c}
    u_{q,l,k}(\rv) \\
    v_{q,l,k}(\rv)
  \end{array}
\right) \ .
\end{equation}
The quasiparticle spinor operators in position and momentum
representations are related by:
\begin{equation}
\Psi(\rv) = \sum_l \int \dd k T_{l,k}(\rv) \Psi_{l,k} \ .
\end{equation}
In the momentum representation, the bare quasiparticle propagator is
of course diagonal:
\begin{equation}\label{GreenMatr}
G_0 (k,l,\omega) = \left(
  \begin{array}{cc}
    \frac{1}{\omega - k + i 0^+} & 0 \\
    0 & \frac{1}{\omega + k - i 0^+}
  \end{array}
\right) \ .
\end{equation}

The momentum representation of the vertex operators is obtained in
the following manner:
\begin{equation}
V_{l_1,k_1;l_2,k_2}^\mu = \int \dd^2 r
  T_{l_1,k_1}^{\dagger}(\rv) V^\mu(\rv) T_{l_2,k_2}^{\phantom{\dagger}}(\rv) \ ,
\end{equation}
where $V^\mu(\rv)$ is either $V_{(1,0),(0,0)}$ or $V_{(0,1),(0,0)}$
depending on $\mu$. When ~(\ref{HMatrEl2}) and ~(\ref{VGF1}) are
substituted in this expression, it is possible to exactly integrate
out the azimuthal angle $\phi$. This implements the physical angular
momentum conservation, so that $V_{l_1,k_1;l_2,k_2}^\mu = 0$ unless
$|l_1-l_2|=1$.

\subsection{The LDOS due to resonant scattering}
\label{sec:plots}

The local density of states is naturally obtained from the
quasiparticle Green's function. Including the quasiparticles at
given position and energy of either spin and angular momentum gives
the following expression in the spinor momentum representation:
\begin{eqnarray}\label{MatrLDOS}
\rho(\epsilon,\rv) & = & -\frac{1}{\pi} \Tr{sign}(\epsilon) \cdot
\Tr{Im} \Biggl\lbrace
   \sum_{l_1,l_2} \int \dd k_1 \dd k_2 \\
& & \Tr{Tr} \Bigl\lbrack
  T_{l_1,k_1}^{\phantom{\dagger}}(\rv) G_{l_1,k_1;l_2,k_2}(\epsilon)
    T_{l_2,k_2}^{\dagger}(\rv)
  \Bigr\rbrack \Biggr\rbrace \ . \nonumber
\end{eqnarray}

In the absence of perturbations, we would substitute here the bare
Green's function $G_0$ from ~(\ref{GreenMatr}). It is easy to show
that the zeroth order term in the LDOS is:
\begin{eqnarray}
&& \rho_0(\epsilon,\rv) = \sum_l \Bigl\lbrace
  \left( |u_{+,l,|\epsilon|}(\rv)|^2+|v_{+,l,|\epsilon|}(\rv)|^2 \right)
  \Theta(\epsilon) \nonumber \\
&& ~~~~~ + \left(
|u_{-,l,|\epsilon|}(\rv)|^2+|v_{-,l,|\epsilon|}(\rv)|^2 \right)
  \Theta(-\epsilon) \Bigr\rbrace \ .
\end{eqnarray}
This is the expected result, which could have been immediately
written from the known eigenfunctions ~(\ref{WF}); it is plotted as
a function of energy in the figure ~\ref{LDOS0}.

The one-loop correction $\rho_1(\epsilon,\rv)$ to LDOS is found by
substituting ~(\ref{G1}) into ~(\ref{MatrLDOS}). The correction to
the Green's function is written in the spinor representation as:
\begin{equation}\label{MatrG1}
G_{l_1,k_1;l_2,k_2}^{(1)}(\omega) =
   G_0(l_1,k_1,\omega) \Sigma_{l_1,k_1;l_2,k_2}(\omega) G_0(l_2,k_2,\omega) \ ,
\end{equation}
where the self-energy matrix is:
\begin{eqnarray}\label{MatrSE}
&& \Sigma_{l_1,k_1;l_2,k_2}(\omega) = \sum_l \int \dd k \\
&& ~~~~~ V^\mu_{l_1,k_1;l,k} \left( \begin{array}{cc}
    \frac{1}{\omega-k-\omega_\Tr{v}+i0^+} & 0 \\
    0 & \frac{1}{\omega+k+\omega_\Tr{v}-i0^+}
  \end{array} \right)
  V^\mu_{l,k;l_2,k_2} \ . \nonumber
\end{eqnarray}
Details of the expression for LDOS and numerical procedures are
elaborated in Appendix ~\ref{appLDOS1}.

\begin{figure}
\includegraphics[width=1.65in]{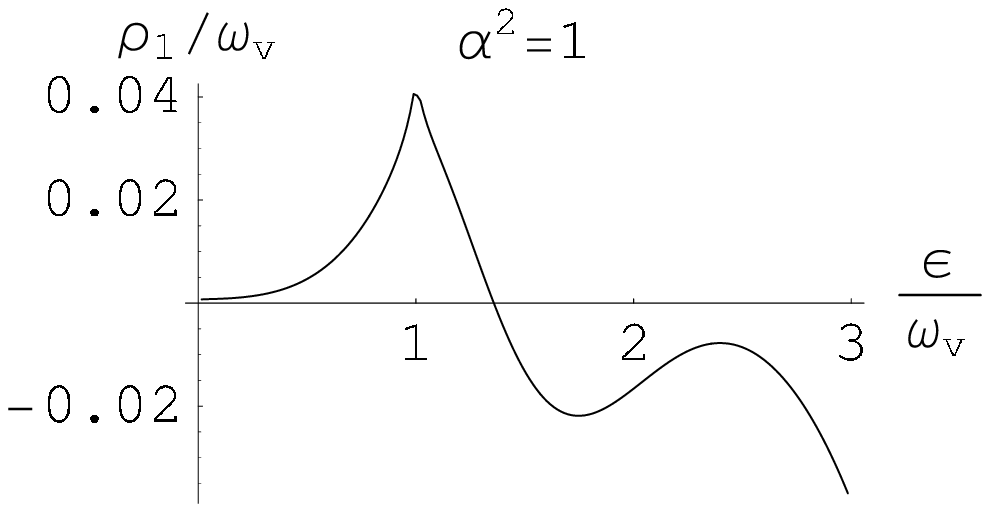}
\includegraphics[width=1.65in]{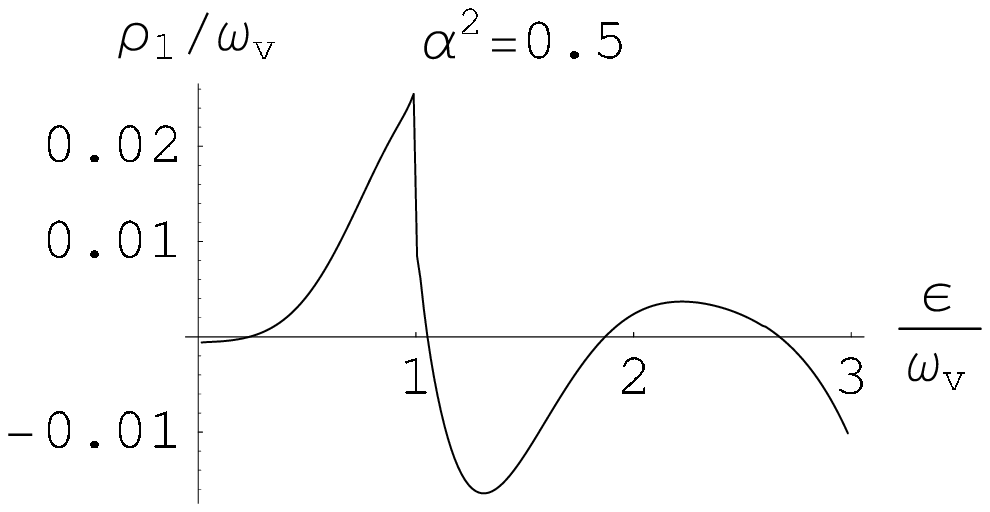}
\includegraphics[width=1.65in]{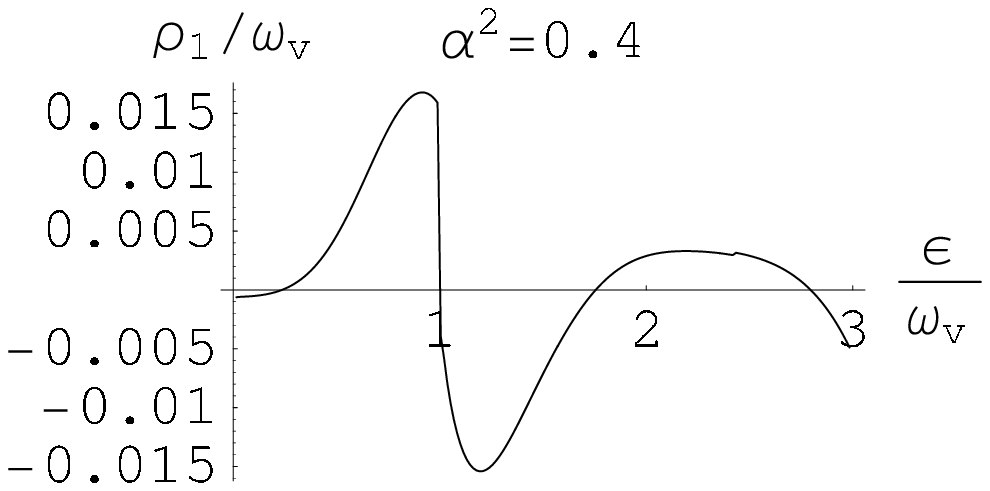}
\includegraphics[width=1.65in]{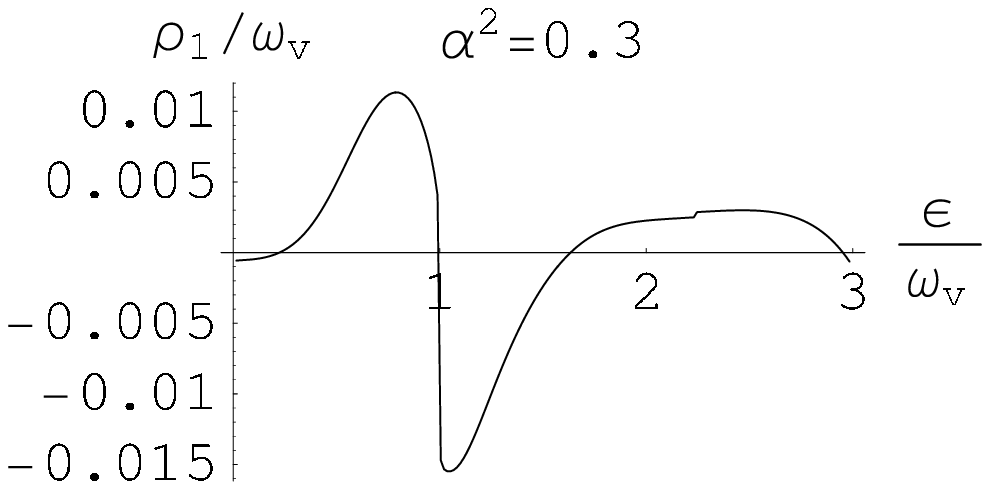}
\includegraphics[width=1.65in]{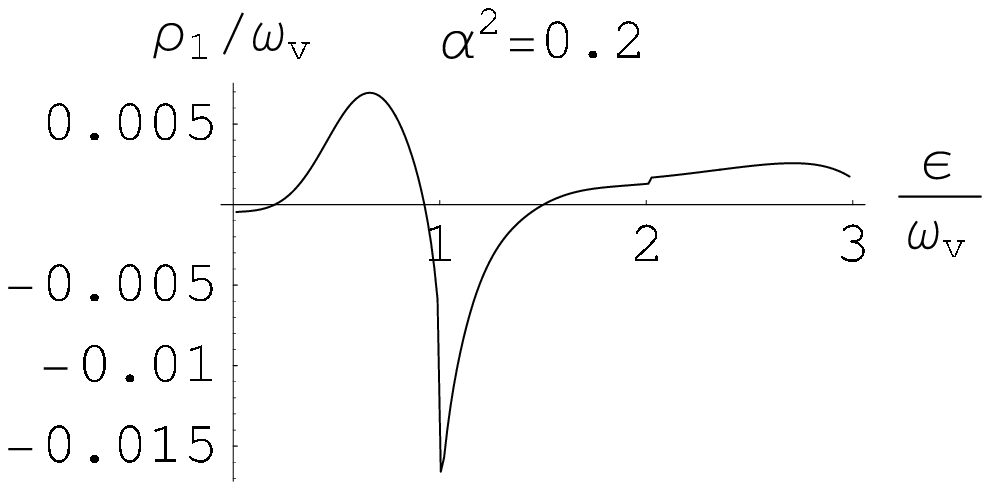}
\includegraphics[width=1.65in]{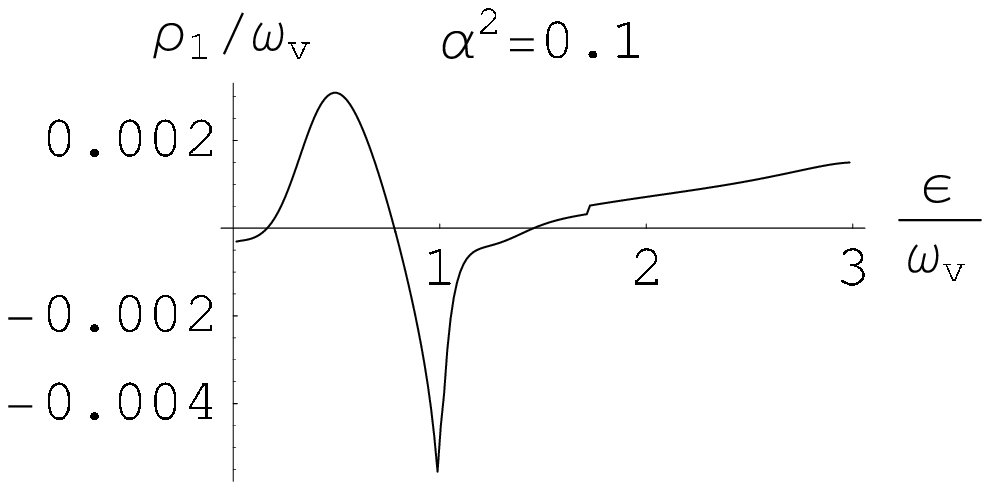}
\includegraphics[width=1.65in]{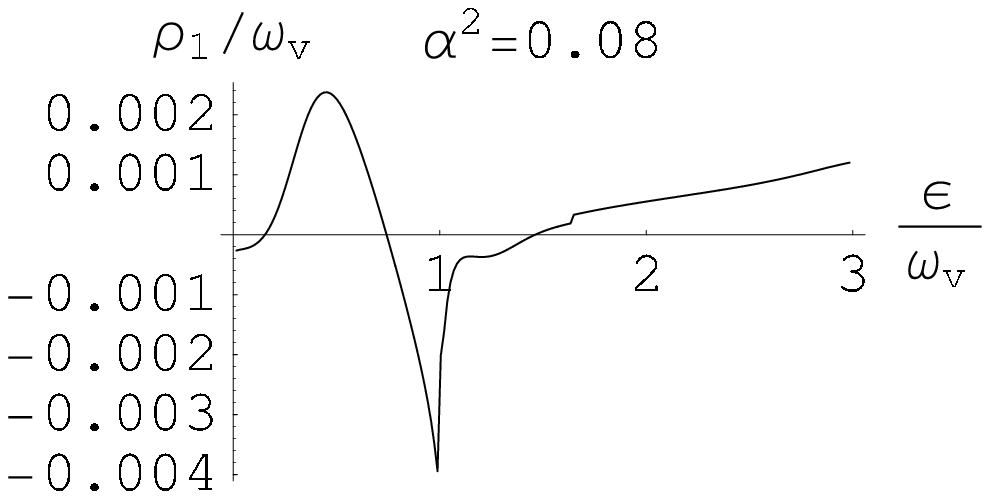}
\includegraphics[width=1.65in]{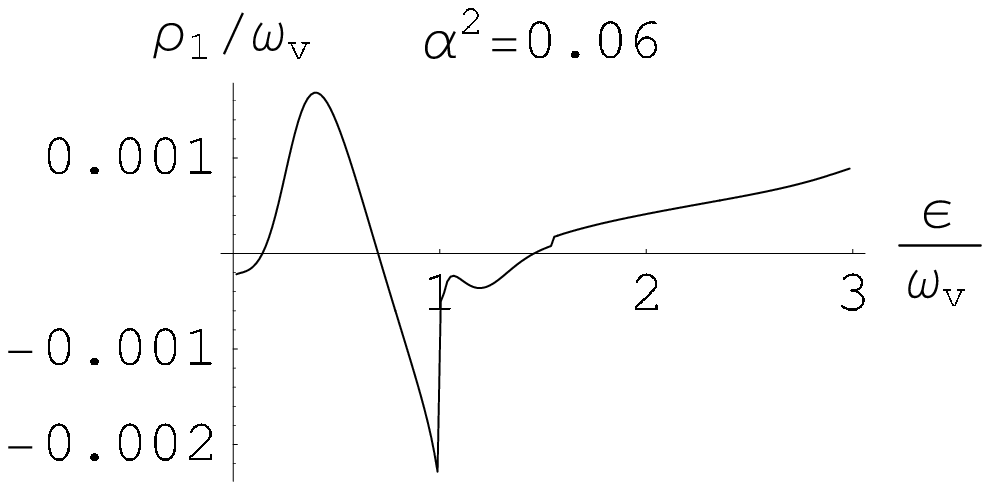}
\includegraphics[width=1.65in]{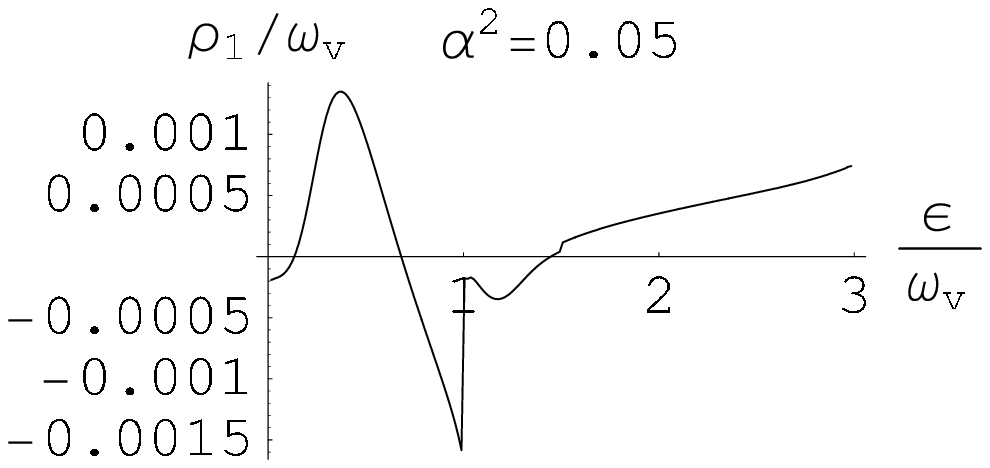}
\includegraphics[width=1.65in]{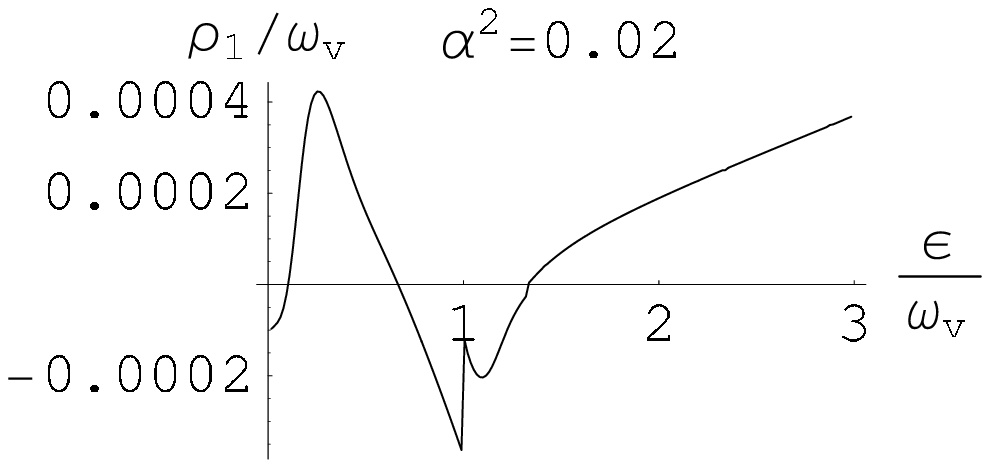}
\includegraphics[width=1.65in]{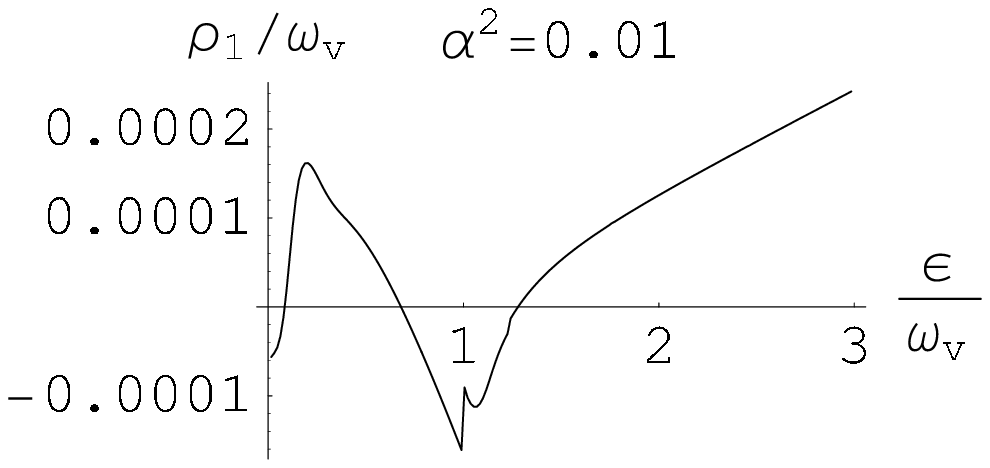}
\includegraphics[width=1.65in]{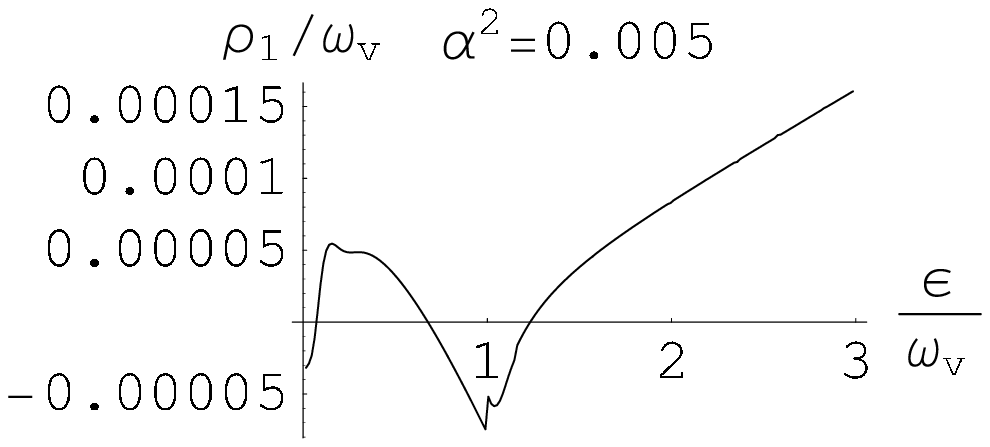}
\caption{\label{LDOS1}The one-loop correction $\rho_1$ to LDOS at
the vortex center as a function of energy. These plots show
evolution of $\rho_1(\epsilon)$ as the small parameter $\alpha^2$
changes. The magnitude of $\rho_1$ scales as $\alpha^2$.
Quasiparticle cut-off energy is $\Lambda=4\omega_\Tr{v}$.}
\end{figure}

The one-loop correction to LDOS at the vortex center is plotted in
the figure ~\ref{LDOS1} as a function of energy. Some general
features can be immediately noted. There is a discontinuity of LDOS
when the quasiparticle energy is equal to the vortex harmonic
frequency, $\epsilon = \omega_\Tr{v}$. Also, for all values of the
small parameter $\alpha^2$, the LDOS has a local minimum at zero
energy and grows into a peak at some finite energy. This feature
could be interpreted as formation of a weak metastable state inside
the spatial region spanned by the vortex quantum oscillations. Note
that the LDOS peak shifts toward lower energies as $\alpha^2 \propto
m_\Tr{v}$ decreases: for smaller vortex mass $m_\Tr{v}$ the vortex
oscillates in a larger region and hence tries to localize
quasiparticles at lower energies. For very small values of
$\alpha^2$ some additional features develop in the LDOS, such as
secondary peaks and dips. It is possible that the values $\alpha^2
\propto 1/n^2$ for integer $n$ are special and control appearance of
various LDOS features (the plots hint to this possibility, but are
not entirely conclusive). Physically, $\alpha$ is the ratio of two
length scales: the quasiparticle wave-length at the energy $\epsilon
= \omega_\Tr{v}$, and the spatial extent of the vortex zero-point
quantum oscillations (effective core radius). Roughly speaking,
features of the LDOS change qualitatively when the effective core
region grows to enclose an additional quasiparticle wavelength $1 /
\omega_\Tr{v}$.

\begin{figure}
\includegraphics[width=2.9in]{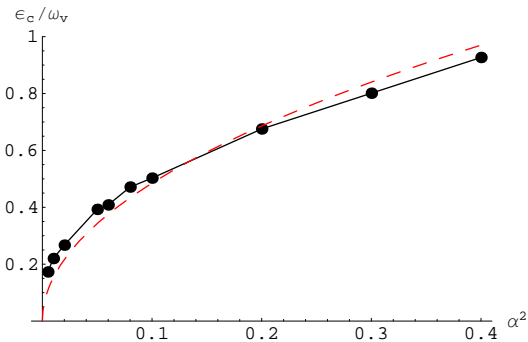}
\caption{\label{PeakEnergy}(color online) Energy $\epsilon_c$ of the
sub-gap peak in the LDOS as a function of the small parameter
$\alpha^2$, and the fit to $\epsilon_c \propto \alpha$ (red dashed).
$\epsilon_c$ is defined as the energy of the first local maximum in
$\rho_1(\epsilon)$ from the figure ~\ref{LDOS1}.}
\end{figure}

In the figure ~\ref{PeakEnergy} we plot energy $\epsilon_c$ of the
LDOS sub-gap peak as a function of $\alpha^2$. The peak position
scales as energy of the perturbation that creates it: $\epsilon_c
\propto \alpha\omega_\Tr{v}$. In order to see how this result
relates to experimental observations, we will explore a somewhat
different scenario. Suppose that the plot in the figure
~\ref{PeakEnergy} were linear, so that $\epsilon_c/\omega_\Tr{v}$
were proportional to $\alpha^2$. This would mean that the sub-gap
peak energy $\epsilon_c$ were proportional to $m_\Tr{v} v_\Tr{f}^2$
(see equation ~(\ref{SmallParameter})). It has been argued
\cite{DWVorAct} that the nodal quasiparticles renormalize the vortex
mass $m_\Tr{v}$ to a value that is proportional to the
superconducting gap amplitude $\Delta_0$. This renormalization due
to quasiparticles is, likely, the largest contribution to the
effective vortex mass, since the usual logarithmic infra-red
divergence of the hydrodynamic vortex mass is cut-off by the Coulomb
screening. Therefore, according to this scenario, $\epsilon_c
\propto \Delta_0$. STM measurements are consistent with such linear
scaling of the sub-gap peak energy with the superconducting gap
amplitude. Most notably, the sub-gap peak position was
experimentally found\cite{aprile,pan,hoogen} not to depend on the
magnetic field, and the considered scenario would agree with this
observation (even though we work with $\omega_\Tr{v}$, which could
depend on the magnetic field through the inter-vortex separation in
a vortex lattice). Discrepency between our actual result
($\epsilon_c \propto \sqrt{\hbar m_\Tr{v} \omega_\Tr{v}
v_\Tr{f}^2}$) and experimental observations could be due to a
combination of large experimental error margins and simplifications
in our model. Further study is needed to appreciate effects of the
Doppler shift and other important factors.

\begin{figure}
\includegraphics[width=2.3in]{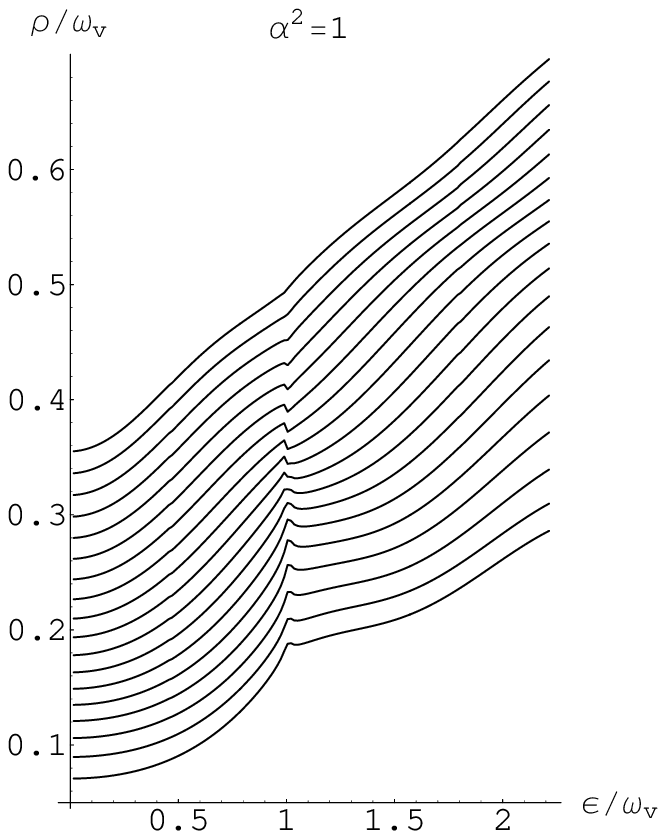}
\includegraphics[width=2.3in]{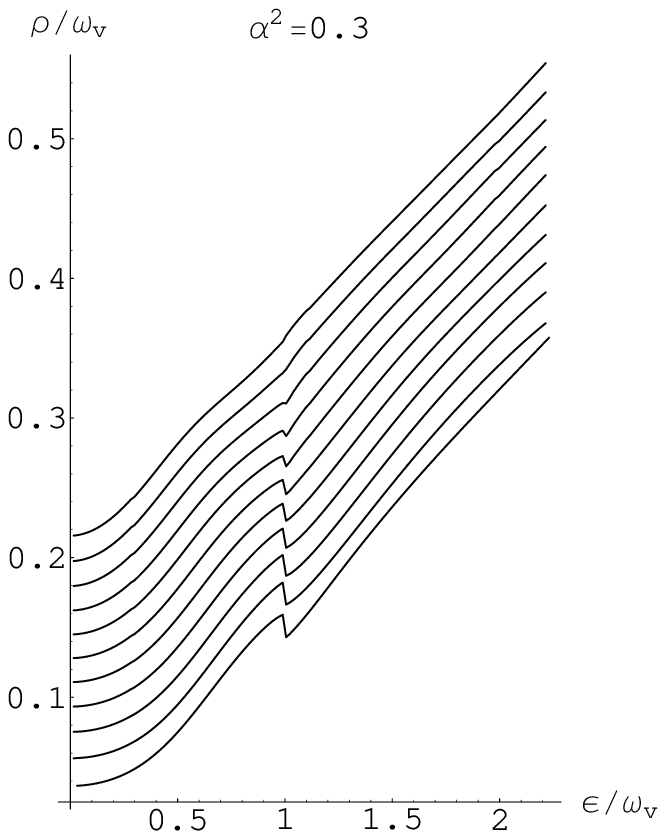}
\caption{\label{LDOStotal}Energy scans of the full LDOS
$\rho=\rho_0+\rho_1$ at gradually increasing distances $r$ from the
vortex core. The plots are offset vertically for clarity, starting
from $r=0$ at the bottom, and moving up with increments $\Delta r =
0.2 \omega_\Tr{v}^{-1}$ for $\alpha^2=1$ and $\Delta r = 0.33
\omega_\Tr{v}^{-1}$ for $\alpha^2 = 0.3$ ($\Delta r
\omega_\Tr{v}\approx (5\alpha)^{-1})$.}
\end{figure}

The full quasiparticle LDOS $\rho = \rho_0 + \rho_1$ as a function
of energy, measured at different distances from the vortex center,
is plotted in the figure ~\ref{LDOStotal}. The sub-gap peak height
is proportional to $\alpha^2$ inside the core, and gradually
decreases over a length-scale comparable to the extent of vortex
zero-point oscillations $(m_\Tr{v} \omega_\Tr{v})^{-1/2}$. Below
$\alpha^2 \approx 0.2$ the sub-gap peak becomes too small to be
visible, just like all other features of the one-loop correction to
LDOS.

Spatial variations in the one-loop correction to LDOS $\rho_1(r)$
are demonstrated in the figure ~\ref{LDOS1vsR} for two values of
$\alpha^2$. In both cases $\rho_1(r)$ was calculated at the energy
of the dominant peak in $\rho_1(\epsilon; r=0)$, which can be seen
in the figure ~\ref{LDOS1}. Correction to the LDOS is largest inside
the region of the vortex quantum oscillations, and hence in
agreement with the conclusion that vortex fluctuations create a
local resonance in the quasiparticle spectrum. All spatial
variations are symmetric under rotations in our model.

\begin{figure}
\includegraphics[width=2.5in]{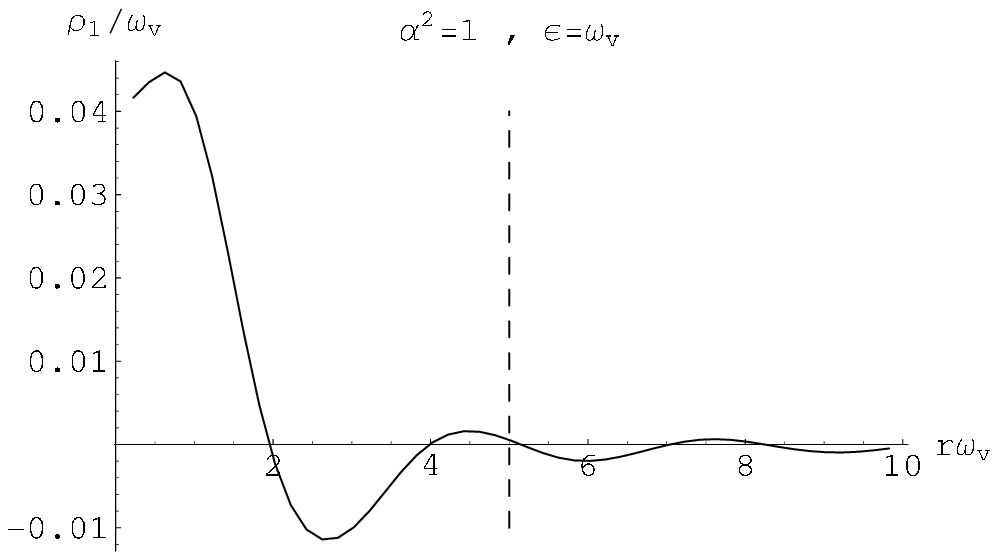}
\includegraphics[width=2.5in]{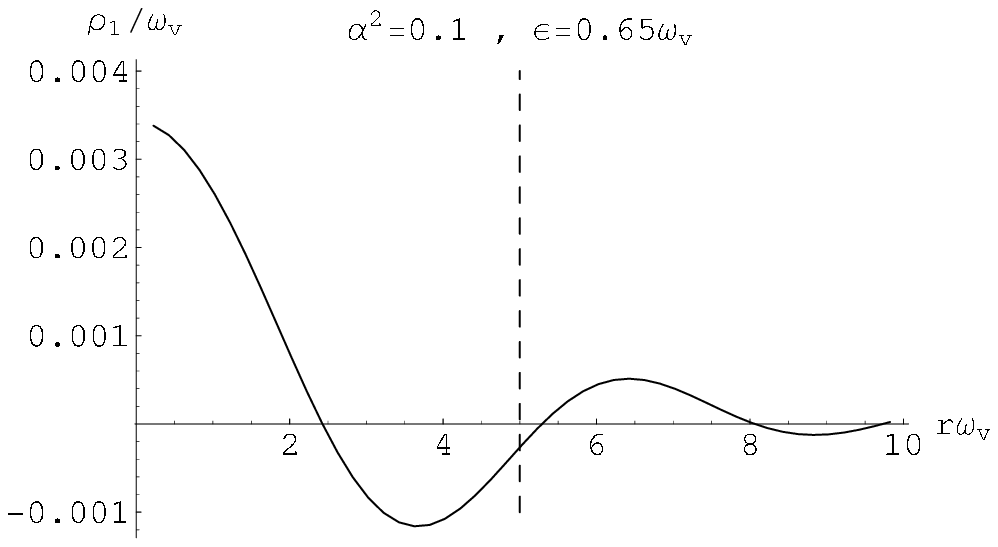}
\caption{\label{LDOS1vsR}The one-loop correction $\rho_1$ of LDOS as
a function of radius. The plots are taken at energies of the
dominant peak in $\rho_1(\epsilon)$. Accuracy is lost to the right
of the dashed line due to the small number of included angular
momentum channels.}
\end{figure}

Finally, in the figure ~\ref{LDOSvsLambda} we explore the influence
of a finite quasiparticle cut-off energy $\Lambda$ on the spectra.
The full LDOS is plotted as function of energy for three values of
$\alpha^2$. The main consequence of finite $\Lambda$ is to reduce
the density of states at large energies. This effect is more
pronounced for large $\alpha^2$ when validity of perturbation theory
becomes questionable. In general, the low energy LDOS is only weakly
affected by $\Lambda$, and the limit $\Lambda \to \infty$ is well
defined.

\begin{figure}
\includegraphics[width=2.1in]{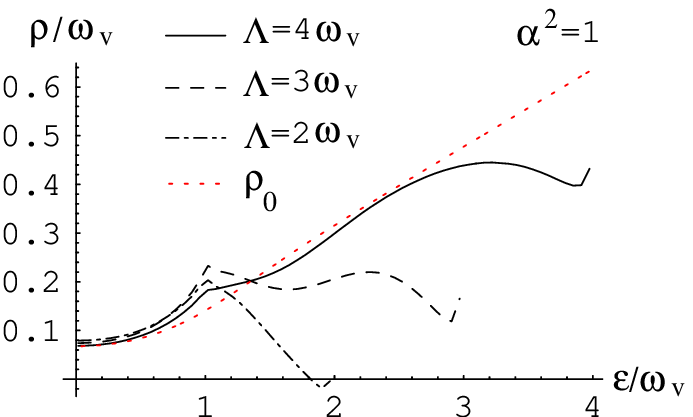}
\vskip 0.2in
\includegraphics[width=2.1in]{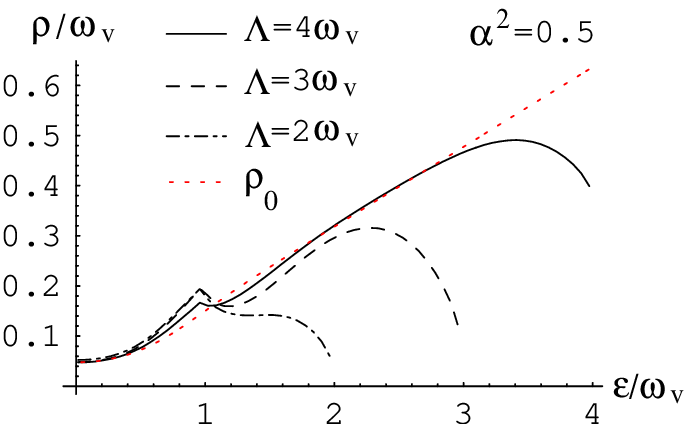}
\vskip 0.2in
\includegraphics[width=2.1in]{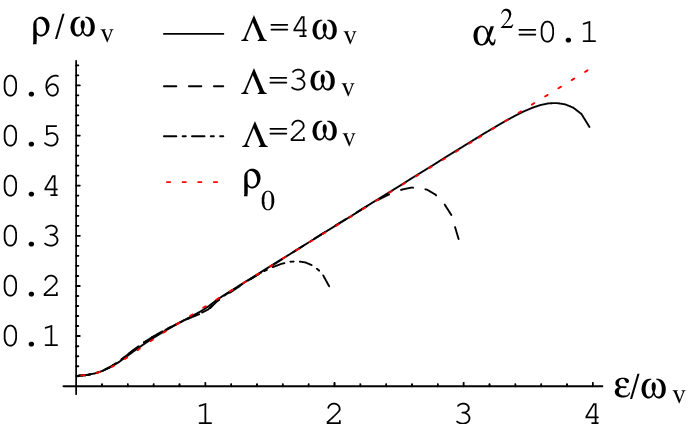}
\caption{\label{LDOSvsLambda}(color online) The full LDOS
$\rho=\rho_0+\rho_1$ at the vortex center as a function of energy
and quasiparticle cut-off $\Lambda$. The dotted red curve is
$\rho_0(\epsilon)$, which excludes the resonant scattering and does
not depend on any finite cut-off.}
\end{figure}

\subsection{Influence of the Magnus force on the LDOS}
\label{sec:magnus}

When a vortex moves with respect to the superfluid, it experiences
the Magus force. This force has the same effect on the vortex as
magnetic field on a moving charged particle, so that it can be
implemented through an effective gauge field $\Bf{\mathcal{A}}$ in
the vortex Hamiltonian:
\begin{equation}\label{VorMagnus}
H_\Tr{v} = \frac{(\pv_\Tr{v}-\Bf{\mathcal{A}})^2}{2m_\Tr{v}} +
\frac{1}{2} m_\Tr{v} \omega_\Tr{v}^2 \rv_\Tr{v}^2 \ ,
\end{equation}
where ($\hbar=1$), for a Galilean-invariant superfluid,
\begin{equation}
\Bf{\nabla} \times \Bf{\mathcal{A}} = 2\pi \rho_\Tr{s} \hat{\Bf{z}}
\ ,
\end{equation}
and $\rho_\Tr{s}$ is the superfluid density (density of Cooper
pairs). On a lattice, recent work \cite{CompOrd1} has argued that
the rhs has to be replaced by difference between the density of
Cooper pairs in the superfluid and half the density  of electrons in
a proximate solid. The presence of the Magnus force defines a
corresponding cyclotron frequency:
\begin{equation}\label{FreqMagnus}
\omega_\Tr{m} = \frac{2\pi\rho_\Tr{s}}{m_\Tr{v}} \ .
\end{equation}
We assume that the vortex trap is isotropic, defining a single trap
harmonic frequency $\omega_\Tr{v}$. Hence, both the trap and the
Magnus force support circular classical trajectories of the vortex,
making it easy to diagonalize the Hamiltonian ~(\ref{VorMagnus}):
\begin{equation}\label{VorMagnus2}
H_\Tr{v} = \omega_0 + \omega_+^{\phantom{\dagger}} a_+^{\dagger}
a_+^{\phantom{\dagger}}
           + \omega_-^{\phantom{\dagger}} a_-^{\dagger} a_-^{\phantom{\dagger}} \ ,
\end{equation}
with
\begin{eqnarray}\label{FreqMagnus2}
\omega_0 & = & \sqrt{\omega_\Tr{v}^2 + \left(\frac{\omega_\Tr{m}}{2}\right)^2} \\
\omega_- & = & \omega_0 - \frac{\omega_\Tr{m}}{2} \nonumber \\
\omega_+ & = & \omega_0 + \frac{\omega_\Tr{m}}{2} \nonumber \ ,
\end{eqnarray}
and
\begin{eqnarray}\label{OpMagnus}
a_- & = & \left(\frac{m_\Tr{v}\omega_0}{2}\right)^{\frac{1}{2}}
          \left( \frac{x_\Tr{v}+iy_\Tr{v}}{\sqrt{2}}
          + \frac{i}{m_\Tr{v}\omega_0} \frac{p_{x\Tr{v}}+ip_{y\Tr{v}}}{\sqrt{2}}
          \right) \\
a_+ & = & \left(\frac{m_\Tr{v}\omega_0}{2}\right)^{\frac{1}{2}}
          \left( \frac{x_\Tr{v}-iy_\Tr{v}}{\sqrt{2}}
          + \frac{i}{m_\Tr{v}\omega_0} \frac{p_{x\Tr{v}}-ip_{y\Tr{v}}}{\sqrt{2}}
          \right) \nonumber \ .
\end{eqnarray}
The full Hamiltonian includes coupling between the vortex and
quasiparticles:
\begin{equation}
H = H_\Tr{v} + \sum_{\Tr{nodes}}
    \int \dd^2 r \Psi^{\dagger}(\rv)  H_{BdG}(\rv) \Psi^{\phantom{\dagger}}(\rv) \ ,
\end{equation}
and we can proceed solving it just like before.

The Magnus force introduces only a few small changes in the LDOS
obtained so far. First, the zero-point quantum motion of the vortex
is now controlled by the frequency $\omega_0$, instead of
$\omega_\Tr{v}$, which can be seen from ~(\ref{OpMagnus}). The
vortex ground-state wavefunction is still isotropic, but extends in
space to distances $\sim (m_\Tr{v} \omega_0)^{-1/2}$ from the trap
center. Second, there are two small parameters, $\alpha_+ =
\sqrt{m_\Tr{v}\omega_0}/\omega_+$ and $\alpha_- =
\sqrt{m_\Tr{v}\omega_0}/\omega_-$ that shape the perturbation
theory. The frequencies $\omega_+$ and $\omega_-$ are important for
the resonant scattering of quasiparticles from the vortex, and in
principle, discontinuities in the LDOS can occur at these
frequencies. The Magnus force lifts the degeneracy of vortex
eigenmodes in an isotropic trap, so that the lowest excited states
correspond to left and right handed circular motion of the vortex,
with different energies.

In the Fig.~\ref{MagnusLDOS} we compare the quasiparticle LDOS for
several values of the Magnus cyclotron frequency $\omega_\Tr{m}$. In
general, no qualitative changes arise when $\omega_\Tr{m}$ is
finite, provided that it does not become too large ($\omega_\Tr{m}
\sim 2\omega_\Tr{v} \Leftrightarrow \alpha_- \to 0$ would invalidate
the perturbation theory). As $\omega_\Tr{m}$ grows, the sub-gap peak
shifts toward smaller energies. Somewhat surprisingly the sharp
features, such as the LDOS discontinuities, seem to become smoother
as $\omega_\Tr{m}$ grows, although they do not completely vanish. In
general, there is only one discontinuity at $\epsilon = \omega_-$.

\begin{figure}
\includegraphics[width=2.1in]{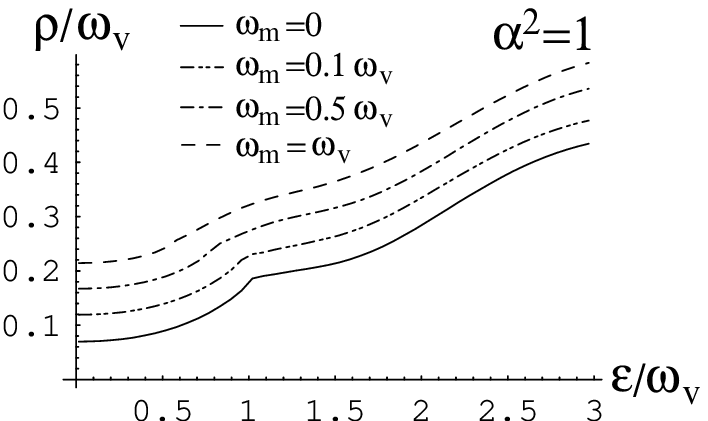}
\vskip 0.2in
\includegraphics[width=2.1in]{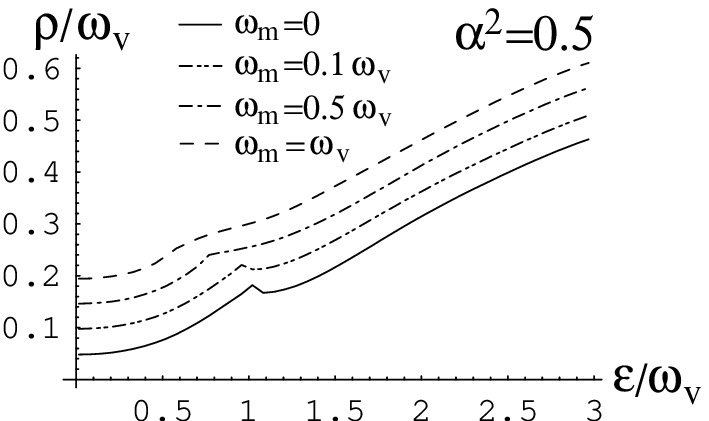}
\vskip 0.2in
\includegraphics[width=2.1in]{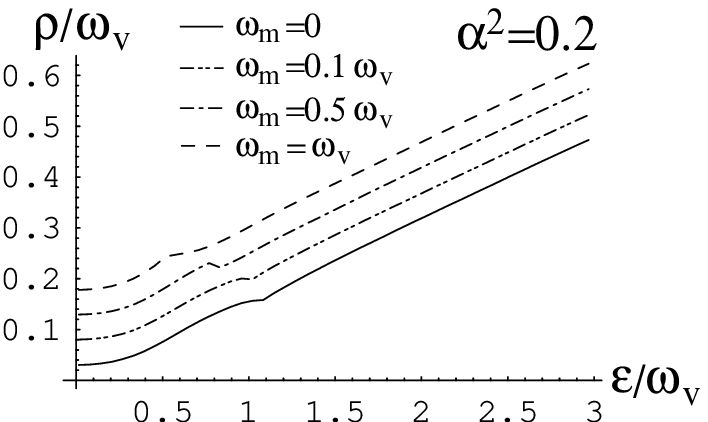}
\caption{\label{MagnusLDOS} The full LDOS at the vortex center as a
function of energy, for different values of the Magnus cyclotron
frequency $\omega_\Tr{m}$ (Eq.~\ref{FreqMagnus}). The plots are
offset vertically for clarity.}
\end{figure}

\section{Discussion and conclusions}
\label{sec:conc}

We have explored how the quantum motion of vortices affects
quasiparticle spectra in clean $d$-wave superconductors at zero
temperature. Keeping only the statistical interaction between
vortices and quasiparticles in a continuum Bogoliubov-de Gennes
model, we have found that quantum oscillations of a vortex could be
responsible for observation of the ``core states'' in the STM
experiments on cuprates. The emerging physical picture is that there
are no bound states in the vortex core of an ideal $d$-wave
superconductor, but quasiparticles can instead experience resonant
scattering from a vortex. Vortices are localized either by their
neighbors in a vortex lattice, or by pinning impurities, and execute
zero-point quantum oscillations in their harmonic traps. When a
quasiparticle scatters from a vortex, it can excite a virtual
higher-energy state of vortex oscillations, and effectively form a
short-lived metastable bound state with the vortex. This leads to a
peak of the local density of states at the appropriate resonant
energy in the vicinity of the vortex core. The peak lies inside the
superconducting gap because the vortex zero-point motion extends
over length-scales that are significantly larger than the coherence
length.

Our numerically calculated LDOS has many similarities with
experimental observations at energies smaller than the
superconducting gap. The low energy scans of LDOS at various
distances from the vortex core look qualitatively the same as the
experimental measurements. We do not obtain a zero-energy peak that
was originally predicted by meanfield BCS calculations
\cite{Wang95}, and we find a small sub-gap peak that gradually
vanishes with growing distance from the vortex core. We also find
other features in the LDOS energy scans, such as discontinuities and
secondary peaks and dips, but these may be blurred and too small to
observe in realistic circumstances. According to our model, energy
of the sub-gap peak turns out to be an increasing, but not a linear
function of the superconducting gap. Experimental data appear to be
consistent with the linear dependence, but due to large error
margins in the measurements, and many simplifications built in our
model, we believe that this detail cannot rule out the great
importance of vortex quantum motion for quasiparticle spectra in
$d$-wave superconductors.

The ability to reach a comprehensive physical picture of the sub-gap
quasiparticle spectra comes with a price: our model contains many
simplifications of realistic circumstances. Most approximations are
controlled in the sense that there is a limit in which they become
accurate. Away from that limit there will be quantitative
modifications of the results due to other effects, such as thermal
fluctuations, decoherence, and possible competing orders inside the
vortex cores. We have also ignored anisotropies of $d$-wave
superconductors. Perhaps the least justified simplification is
inclusion of only the statistical interactions between vortices and
quasiparticles. The missing Doppler shift presents an effective
potential to quasiparticles near the vortex core, and may enhance
the resonant scattering. We do not expect that it could
qualitatively change the results, but its quantitative contribution
need not be small.

We also reiterate here the original motivation for considering the
vortex zero-point motion model. This arose from its consequences in
the {\em spatial\/} dependence of the LDOS: it was argued
\cite{CompOrd1,VorMass} that the Aharanov-Bohm phases acquired by
vortices from the background density of electrons lead naturally to
periodic modulations in the LDOS. This was proposed as an
explanation for the modulations observed in STM experiments
\cite{hoffman,fischer}. Here we have shown that the same model can
also help account for some crucial aspects of the {\em energy\/}
dependence of the LDOS. Further, as argued in
Ref.~\onlinecite{VorMass}, there is also a quantitative consistency
between the energy scales deduced from the spatial and energy
dependencies. From the extent of the spatial
modulations,\cite{hoffman} and an estimate of the trapping potential
on a vortex, we were able to obtain an estimate of the vortex
oscillation frequency.\cite{VorMass} This number is consistent with
the observed position\cite{aprile,pan,hoffman} of the sub-gap peaks
in the energy dependence of the LDOS.

Future experiments may provide more ways to detect consequences of
vortex quantum motion in $d$-wave superconductors. Neutron or light
scattering measurements could provide a direct observation of the
vortex oscillations. The spatial extent of the LDOS modulations
provides means for measuring one important energy scale,
$\sqrt{\hbar \omega_\Tr{v} m_\Tr{v} v_\Tr{f}^2}$, where
$\omega_\Tr{v}$ is the harmonic frequency of a localized vortex. Our
present prediction is that the sub-gap peak in the quasiparticle
LDOS occurs at the energy proportional to this scale. An increased
sensitivity of STM measurements in the future could allow resolving
the jumps in the LDOS at energies that reflect the discrete spectrum
of the trapped quantum vortex. This would provide means for
measuring another important energy scale, $\hbar \omega_\Tr{v}$.
Combined knowledge of both energy scales reveals the vortex mass
$m_\Tr{v}$ and the strength of the vortex trapping force
(inter-vortex interactions), which in turn can be related to other
parameters, such as the superconducting gap and magnetic field.
Effects of the Magnus force reduce to quantitative modifications of
these energy scales, as discussed in the Section~\ref{sec:magnus}.

\acknowledgments

This research was supported by the National Science Foundation under
grant DMR-0537077.

\appendix

\section{Calculation of vertex operators}\label{appVertex}

Here we derive general expressions for the ``matrix elements''
~(\ref{HMatrEl}), which play the role of vertex operators in the
perturbation theory. The matrix elements are evaluated by
substituting the wavefunctions of the linear harmonic oscillator
into ~(\ref{HMatrEl}), and have the same form as the Hamiltonian
~(\ref{IsoBdG}) in which the vector operator $\Bf{a}$ is substituted
by the vector c-number:
\begin{eqnarray}\label{aeff}
&& \Bf{a}_{(n_{1x},n_{1y}),(n_{2x},n_{2y})}(\rv) = \frac{m_\Tr{v}
\omega_\Tr{v}}{2\pi}
   \frac{2^{-(n_{1x}+n_{1y}+n_{2x}+n_{2y})/2}}{\sqrt{n_{1x}! n_{2x}! n_{1y}! n_{2y}!}}
   \nonumber \\
&& ~~~ \times \int \dd^2 r_\Tr{v} \frac{\hat{\Bf{z}} \times (\rv -
\rv_\Tr{v})}{2 |\rv
       - \rv_\Tr{v}|^2} e^{-m_\Tr{v} \omega_\Tr{v} r_\Tr{v}^2} \\
&& ~~~ \times H_{n_{1x}}(x_\Tr{v} \sqrt{m_\Tr{v} \omega_\Tr{v}})
              H_{n_{1y}}(y_\Tr{v} \sqrt{m_\Tr{v} \omega_\Tr{v}}) \nonumber \\
&& ~~~ \times H_{n_{2x}}(x_\Tr{v} \sqrt{m_\Tr{v} \omega_\Tr{v}})
              H_{n_{2y}}(y_\Tr{v} \sqrt{m_\Tr{v} \omega_\Tr{v}}) \ . \nonumber
\end{eqnarray}
$H_n(x)$ are Hermite polynomials of order $n$. One way to
analytically calculate these integrals is to use the following
formula:
\begin{equation}
\frac{\hat{\Bf{z}} \times (\rv - \rv_\Tr{v})}{2 |\rv -
\rv_\Tr{v}|^2} = \frac{1}{2} \hat{\Bf{z}} \times \left( \Bf{\nabla}
\log | \rv - \rv_\Tr{v} | \right) \ ,
\end{equation}
and expand:
\begin{eqnarray}
&& \log|\rv-\rv_\Tr{v}| = \log|z-z_\Tr{v}|
  = \log \left\vert z_> \left( 1 - \frac{z_<}{z_>} \right) \right\vert \nonumber \\
&& ~~~~~ = \log |z_>| + \frac{1}{2}
\log\left(1-\frac{z_<}{z_>}\right) + \frac{1}{2}
           \log\left(1-\frac{z_<^*}{z_>^*}\right) \nonumber \\
&& ~~~~~ = \log |z_>|  - \Tr{Re} \sum_{n=1}^\infty \frac{1}{n}
\left( \frac{z_<}{z_>}
           \right)^n \ .
\end{eqnarray}
Coordinates have been represented by complex numbers $z$ and
$z_\Tr{v}$; $z_>$ and $z_<$ are the greater and lesser by modulus of
$z$ and $z_\Tr{v}$ respectively. The effective gauge-field
~(\ref{aeff}) is:
\begin{eqnarray}
&& \Bf{a}_{(n_{1x},n_{1y}),(n_{2x},n_{2y})}(\rv) ~ \propto ~ \\
&& ~~~ \propto \hat{\Bf{z}} \times \Bf{\nabla} \int \dd^2 z_\Tr{v}
       \left\lbrack \log |z_>|  - \Tr{Re} \sum_{n=1}^\infty \frac{1}{n}
       \left( \frac{z_<}{z_>} \right)^n \right\rbrack \times \nonumber \\
&& ~~~~~~~ e^{-m_\Tr{v} \omega_\Tr{v} |z_\Tr{v}|^2}
       H_{n_{1x}} H_{n_{1y}} H_{n_{2x}} H_{n_{2y}} \ . \nonumber
\end{eqnarray}
The Hermite polynomials are easily written in terms of $z_\Tr{v}$
and $z_\Tr{v}^*$ at low orders $n$. For any particular set of
$(n_{1x},n_{1y},n_{2x},n_{2y})$ only one term in the expansion above
gives a non-zero contribution upon integration of the phase of
$z_\Tr{v}$. For example, when $n_{1x} = n_{1y} = n_{2x} =n_{2y} = 0$
all Hermite polynomials are equal to unity, and the only term that
does not depend on phase is the logarithm:
\begin{eqnarray}
\Bf{a}_{(0,0),(0,0)} & = & m_\Tr{v} \omega_\Tr{v} \hat{\Bf{z}}
\times \Bf{\nabla}
     \Biggl\lbrack \int\limits_0^{|z|} \dd |z_\Tr{v}| e^{-m_\Tr{v} \omega_\Tr{v}
       |z_\Tr{v}|^2} \log(|z|) + \nonumber \\
& &  \int\limits_{|z|}^{\infty} \dd |z_\Tr{v}| e^{-m_\Tr{v}
\omega_\Tr{v}
       |z_\Tr{v}|^2} \log(|z_\Tr{v}|) \Biggr\rbrack \\
& = & m_\Tr{v} \omega_\Tr{v} \hat{\Bf{z}} \times \hat{\Bf{r}}
       \frac{\partial}{\partial r} \log(r)
      \int_0^r \dd r_\Tr{v} r_\Tr{v} e^{-m_\Tr{v} \omega_\Tr{v} r_\Tr{v}^2} \nonumber \\
& = & \hat{\Bf{z}} \times \frac{\Bf{r}}{r^2}
       \left( 1 - e^{-m_\Tr{v} \omega_\Tr{v} r^2} \right) \ . \nonumber
\end{eqnarray}
This is the result written in ~(\ref{VGF}). A similar calculation
produces the expression ~(\ref{VGF1}).

\section{Calculation of the one-loop correction to the LDOS}\label{appLDOS1}

Here we substitute ~(\ref{G1}) and ~(\ref{SEfock}) into
~(\ref{MatrLDOS}) and derive the lowest order correction to the
LDOS. We systematically use the formula:
\begin{equation}
\frac{1}{\omega-\epsilon+i0^+} = \frac{\mathbb{P}}{\omega-\epsilon}
- i\pi\delta(\omega-\epsilon)
\end{equation}
to extract the imaginary part in ~(\ref{MatrLDOS}). After a few
algebraic manipulations one obtains the following correction to the
LDOS:
\begin{widetext}
\begin{eqnarray}\label{LDOSoneloop}
& & \rho_1(\epsilon,\rv)
= \Tr{sign}(\epsilon) \sum_{l_1,l_2} \sum_l \Tr{Tr} \Biggl\lbrace \\
& & -\pi^2
   T_{l_1,|\epsilon|}^{\phantom{\dagger}}(\rv)
   \left( \begin{array}{cc} \ss \Theta(\epsilon) & \ss 0 \\ \ss 0 & \ss -\Theta(-\epsilon) \end{array} \right)
   V^\mu_{l_1,|\epsilon|;l,|\epsilon|-\omega_\Tr{v}}
   \left( \begin{array}{cc} \ss \Theta(\epsilon-\omega_\Tr{v}) & \ss 0 \\ \ss 0 & \ss -\Theta(-\epsilon-\omega_\Tr{v}) \end{array} \right)
   V^\mu_{l,|\epsilon|-\omega_\Tr{v};l_2,|\epsilon|}
   \left( \begin{array}{cc} \ss \Theta(\epsilon) & \ss 0 \\ \ss 0 & \ss -\Theta(-\epsilon) \end{array} \right)
   T_{l_2,|\epsilon|}^{\dagger}(\rv)
\nonumber \\ & & + \int \dd k_1 \dd k_2
   T_{l_1,k_1}^{\phantom{\dagger}}(\rv)
   \left( \begin{array}{cc} \ss \frac{\mathbb{P}}{\epsilon-k_1} & \ss 0 \\ \ss 0 & \ss \frac{\mathbb{P}}{\epsilon+k_1} \end{array} \right)
   V^\mu_{l_1,k_1;l,|\epsilon|-\omega_\Tr{v}}
   \left( \begin{array}{cc} \ss \Theta(\epsilon-\omega_\Tr{v}) & \ss 0 \\ \ss 0 & \ss -\Theta(-\epsilon-\omega_\Tr{v}) \end{array} \right)
   V^\mu_{l,|\epsilon|-\omega_\Tr{v};l_2,k_2}
   \left( \begin{array}{cc} \ss \frac{\mathbb{P}}{\epsilon-k_2} & \ss 0 \\ \ss 0 & \ss \frac{\mathbb{P}}{\epsilon+k_2} \end{array} \right)
   T_{l_2,k_2}^{\dagger}(\rv)
\nonumber \\ & & + \int \dd k_2 \dd k
   T_{l_1,|\epsilon|}^{\phantom{\dagger}}(\rv)
   \left( \begin{array}{cc} \ss \Theta(\epsilon) & \ss 0 \\ \ss 0 & \ss -\Theta(-\epsilon) \end{array} \right)
   V^\mu_{l_1,|\epsilon|;l,k}
   \left( \begin{array}{cc} \ss \frac{\mathbb{P}}{\epsilon-k-\omega_\Tr{v}} & \ss 0 \\ \ss 0 & \ss \frac{\mathbb{P}}{\epsilon+k+\omega_\Tr{v}} \end{array} \right)
   V^\mu_{l,k;l_2,k_2}
   \left( \begin{array}{cc} \ss \frac{\mathbb{P}}{\epsilon-k_2} & \ss 0 \\ \ss 0 & \ss \frac{\mathbb{P}}{\epsilon+k_2} \end{array} \right)
   T_{l_2,k_2}^{\dagger}(\rv)
\nonumber \\ & & + \int \dd k \dd k_1
   T_{l_1,k_1}^{\phantom{\dagger}}(\rv)
   \left( \begin{array}{cc} \ss \frac{\mathbb{P}}{\epsilon-k_1} & \ss 0 \\ \ss 0 & \ss \frac{\mathbb{P}}{\epsilon+k_1} \end{array} \right)
   V^\mu_{l_1,k_1;l,k}
   \left( \begin{array}{cc} \ss \frac{\mathbb{P}}{\epsilon-k-\omega_\Tr{v}} & \ss 0 \\ \ss 0 & \ss \frac{\mathbb{P}}{\epsilon+k+\omega_\Tr{v}} \end{array} \right)
   V^\mu_{l,k;l_2,|\epsilon|}
   \left( \begin{array}{cc} \ss \Theta(\epsilon) & \ss 0 \\ \ss 0 & \ss -\Theta(-\epsilon) \end{array} \right)
   T_{l_2,|\epsilon|}^{\dagger}(\rv)
\Biggr\rbrace \ . \nonumber
\end{eqnarray}
\end{widetext}

This expression for $\rho_1$ is written assuming that the sample is
infinite. In order to carry out numerical calculations it is
necessary to impose a finite sample radius $R$ and thus quantize
energy levels of the unperturbed Hamiltonian. The quantized energies
take values $\epsilon_n = \pm k_n$, where $k_n R + \delta$ is the
$n^\Tr{th}$ zero of the Bessel function ($\delta$ is a phase shift).
In the region $k_n R \gg 1$, the energy levels are separated by
$\pi/R$, and this allows a simple rule of thumb for conversion
between the continuous and discrete computations. First, the
infinite-sample wavefunctions need to be renormalized to unity on
the finite sample. Up to a small error of the order $(k_n R)^{-1}$
this amounts to multiplying the wavefunction by $\sqrt{\pi / R}$.
Second, all integrations over continuous radial wave-vectors $k$ in
the expression ~(\ref{LDOSoneloop}) must be replaced by summations
over discrete $k_n$, with a measure factor of $\pi / R$. Finally,
the principal part $\mathbb{P}/x$ is taken to deviate from $1/x$ in
the interval of width $\pi / R$ from $x=0$, passing smoothly through
zero at $x=0$.

As a result of energy discretization, the LDOS is strictly speaking
defined only at discrete energy levels $\epsilon_n$. Such
discretization is naturally lifted if there are additional
fluctuations in the problem that broaden the energy levels beyond
$\pi / R$ (for example, due to finite temperature or disorder).
However, the level-broadening mechanisms are not important in the $R
\to \infty$ limit, which is reached in numerics if the majority of
states satisfy $k_n R \gg 1$ (that is, if the energy cut-off
$\Lambda$ is large, $\Lambda R \gg 1$). The typical values that we
used in numerics were $R=200$, $\Lambda=4$, $\omega_\Tr{v}=1$, and
$-5\leq l \leq 5$.

\section{Microscopic theory for a $d$-wave superconductor}
\label{paper1}

This appendix connects the results of the present paper to those of
paper I. Here we will extend the numerical results of I to $d$-wave
superconductors. However, because of the gradient expansion
necessary in this method, we are unable to extend these results to
very small vortex core sizes. Consequently, we will not find an
elimination of the zero bias peak in the LDOS.

As discussed in I, we will use the gap operator given in Eq.~(3) of
I, with $\Delta({\bf r}) = \Delta(r) e^{i\theta}$. In polar
coordinates which are more convenient for our problem here this gap
operator reads
\begin{widetext}
  \begin{align}
    \hat \Delta =
    & \frac{e^{3i\theta}}{8 k_F^2 r^2} \left[
        (3\Delta - 3 r \Delta^{\prime} + r^2 \Delta^{\prime\prime})
        +4(-2 r \Delta + r^2 \Delta^{\prime}) \partial_r
        +4i(-3 \Delta + r \Delta^{\prime}) \partial_{\theta}
        +4\Delta(r^2 \partial_r^2 - \partial_{\theta}^2
        + 2ir\partial_r \partial_{\theta}) \right] \nonumber \\
    +  &
    \frac{e^{-i\theta}}{8 k_F^2 r^2} \left[
        (-\Delta + r \Delta^{\prime} + r^2 \Delta^{\prime\prime})
        +4 r^2 \Delta^{\prime} \partial_r
        +4i(\Delta - r \Delta^{\prime}) \partial_{\theta}
        +4\Delta(r^2 \partial_r^2 - \partial_{\theta}^2
        - 2ir\partial_r \partial_{\theta}) \right] \;.
    \label{eq:polargapoperator}
  \end{align}
\end{widetext}
As a generalization of Eqs.~(21) and (31) of I
we write the quasi-particle wave functions
as
\begin{equation}
  \left(\begin{array}{c}
      u_{\ell} ({\bf r}) \\ v_{\ell} ({\bf r})
        \end{array}\right) = \sum_{m j}
        \left(\begin{array}{c}
           c_{m j}^{(\ell)} \phi_{m,j} ({\bf r}) \\
           d_{m j}^{(\ell)} \phi_{m,j} ({\bf r})
        \end{array}\right) \;,
\label{eq:ufd}
\end{equation}
where the
\begin{equation}
\phi_{m,j} ({\bf r}) \equiv \phi_{m,j}(r,\theta) \equiv \exp[-i m
\theta] \phi_{m j}(r)/\sqrt{2\pi}
\end{equation}
[which generalize Eq.~(29) of I] now form a complete set of
eigenfunctions to the kinetic energy operator and satisfy the
normalization condition
\begin{equation}
  \label{eq:normalizationd}
  \int_0^{2\pi} d\theta \int_0^{R_0} dr\, r \,
  \phi_{m j}^{\ast}(r,\theta) \, \phi_{m' j'}(r,\theta) = \delta_{m m'} \delta_{j j'} \;.
\end{equation}
The Bogoliubov-de Gennes equations reduce to a matrix equation quite
similar to Eq.~(32) of I. 
The essential difference is
that there is now an infinite hierarchy of coupled angular momentum
channels and the matrices $T^{\pm}$ and $\Delta$ now have matrix
elements
\begin{align}
  \label{eq:Td}
  T_{m j;m' j'}^{\pm} = & \, \mp \frac{1}{2m_e} \left(
    \frac{\alpha_{m,j}^2}{R_0^2} - k_F^2 \right) \delta_{m,m'} \delta_{j j'}\;,  \\
  \label{eq:Deltad}
  \Delta_{m j; m' j'} = & \int_0^{2\pi} d\theta \int_0^{R_0} dr\, r \, \phi_{m,j}^{\ast}(r,\theta) 
\hat \Delta\, \phi_{m',j'}(r,\theta) \;.
\end{align}
Using the gap operator given in Eq.~(\ref{eq:polargapoperator}) we
can explicitly evaluate $\Delta_{m j; m' j'}$ and obtain
\begin{widetext}
  \begin{align}
    \Delta_{m j; m' j'} = \delta_{m',m+3} &
    \int_0^{R_0} dr\, \frac{\phi_{m,j}(r)}{8 k_F^2 r}
         \big\{ \left[ (-4(\alpha_{m+3,j'}r/R_0)^2 + 3) \Delta
        -3 r \Delta' + r^2 \Delta'' \right] \phi_{m+3,j'} (r)  \big. \nonumber \\
        \big. & \qquad \qquad \qquad \qquad \quad {} +  4 \left[(2m+3) \Delta + r \Delta' \right] \phi_{m+3,j'}^{-} (r)
        \big\} \nonumber \\
        {} +\delta_{m',m-1} &
    \int        _0^{R_0} dr\, \frac{\phi_{m,j}(r)}{8 k_F^2 r}
        \big\{ \left[ (-4(\alpha_{m-1,j'}r/R_0)^2 -1) \Delta
        + r \Delta' + r^2 \Delta'' \right] \phi_{m-1,j'} (r)  \big. \nonumber \\
        \big. & \qquad \qquad \qquad \qquad \quad {} +  4 \left[(2m-1) \Delta - r \Delta' \right] \phi_{m-1,j'}^{+} (r)
        \big\} \;.
         \label{eq:dDeltaII}
  \end{align}
\end{widetext}
The functions
\begin{equation}
\phi_{m,j}^{\pm} \equiv \frac{\sqrt{2}\, \alpha_{m,j} r}{R_0^2
|J_{m+1}(\alpha_{m,j})|} J_{m \pm 1}(\alpha_{m,j}r/R_0)
\end{equation}
 are modifications of the normalized Bessel functions $\phi_{m,j}(r)$ defined in Eq.~(29) of I 
which arise when taking derivatives of $\phi_{m,j}(r)$.
According to Eq.~(\ref{eq:dDeltaII}) $m'$ has to equal $m-1$ or
$m+3$ for a matrix element not to vanish. The reason for this is
that while the vortex in the order parameter mediates a change in
$m$ (minus angular momentum) by $1$ when scattering the hole-like
component of the quasi-particle into an electron-like component, due
to the $d$-wave symmetry of the gap operator we have an additional
change by $\pm 2$. As a consequence of this, the eigenvalue problem
breaks up into four apparently independent eigenvalue problems which
we all truncate at large angular momentum $m_{\text{max}}$.

Let us say that the eigenvector $\Psi_{\ell}$ lies in the sector
$s(\ell)=0,1,2$, or $3$ if $\Psi_{\ell}$ contains only components
$c_{4\nu+s}$ and $d_{4\nu+s-1}$ with $\nu$ an integer. It is
possible to write the eigenvalue problem for each of these four
sectors in terms of a real and symmetric band-diagonal matrix. For
concreteness, let us consider the sector $s=0$ with $m_{\text{max}}
= 5$ (which, of course, for practical calculations should be chosen
much larger). We then have
\begin{widetext}
  \begin{equation}
    \left( \begin{array}{llllll}
        T_{-5}^{+} & \Delta_{-4;-5}^T & & & & 0 \\
        \Delta_{-4;-5} & T_{-4}^{-} & \Delta_{-4;-1} & & & \\
        & \Delta_{-4;-1}^T & T_{-1}^{+} & \Delta_{0;-1}^T & & \\
        & & \Delta_{0;-1} & T_{0}^{-} & \Delta_{0;3} & \\
        & & & \Delta_{0;3}^T & T_{3}^{+} & \Delta_{4;3}^T \\
        0 & & & & \Delta_{4;3} & T_{4}^{-} \\
      \end{array} \right) \Psi_{\ell}^{(s)} = \epsilon_{\ell}^{(s)}
    \,      \Psi_{\ell}^{(s)} \;.
  \end{equation}
\end{widetext}
Here,
 $T_{m}^{\pm}$ and $\Delta_{m;m'}$ are themselves $N_0 \times N_0$ matrices whose matrix elements are given above.
We can now relate each eigenstate in sector $1$ to one in sector $0$
and each eigenstate in sector $3$ to one in sector $2$ by making use
of the property that if $[u_\ell({\bf r}),v_\ell({\bf r})]^T$ is an
eigenstate of $\mathcal{H}_{\text{BdG}}^0$ with eigenvalue
$\epsilon_\ell$, then $[-v_\ell^{\ast}({\bf r}),u_\ell^{\ast}({\bf
r})]^T$ is an eigenstate of $\mathcal{H}_{\text{BdG}}^0$ with
eigenvalue $-\epsilon_\ell$. This property implies the symmetry
\begin{equation}
  \Delta_{m j; m' j'} = -\Delta_{-m', j'; -m, j}\;,
\end{equation}
which we have checked explicitly.\footnote{In the more general case
there is an extra factor $-(-1)^{m-m'}$ which for the $d$-wave case
considered here equals one for all non-zero matrix elements.} In
terms of the eigenvectors $\Psi = (c_{4\nu+s},d_{4\nu+s-1})$  we
find that if $\Psi$ is an eigenvector in sector $s$ with eigenvalue
$\epsilon$, then $\tilde \Psi = (\tilde c_{4\nu+(1-s)},\tilde
d_{4\nu+(1-s)-1})$ with $\tilde c_{m,j} \equiv d_{-m,j}$ and $\tilde
d_{m,j} \equiv c_{-m,j}$ is an eigenvector with eigenvalue
$-\epsilon$ in sector $1-s\,(\text{mod}\, 4)$. It is therefore
sufficient to only consider the sectors $0$ and $2$ and make use of
this property to obtain all eigenstates.

An interesting property of having no mixing between the four sectors
is that the quasi particle amplitudes $u_{\ell}^{(s)}({\bf r})$ and
$v_{\ell}^{(s)}({\bf r})$ can be written as
\begin{align}
  u_{\ell}^{(s)} ({\bf r}) = & \frac{e^{-i s \theta}}{\sqrt{2\pi}} \sum_{\nu,j} c_{4\nu+s,j}^{(\ell)}
  e^{-4i\nu \theta} \phi_{4\nu+s,j}(r) \;,   \label{eq:u4} \\
  v_{\ell}^{(s)} ({\bf r}) = & \frac{e^{-i (s-1) \theta}}{\sqrt{2\pi}} \sum_{\nu,j} d_{4\nu+s-1,j}^{(\ell)} e^{-4i\nu \theta} \phi_{4\nu+s-1,j}(r) \;.
\end{align}
Physically measurable quantities like the local DOS which only
involve $|u_{\ell}^{(s)}({\bf r})|^2$ (and $|v_{\ell}^{(s)}({\bf
r})|^2$) are therefore invariant under all symmetry operations
expected for a $d_{x^2-y^2}$-wave order parameter, i.e.\ rotations
by $90^{\circ}$ as well as reflections along the nodal or anti-nodal
directions. Before considering the effect of the zero-point
fluctuations of the vortex on the local DOS let us first have a look
at the case of a static vortex which was already studied in Refs.
\onlinecite{Wang95,Franz98}. One advantage of using the gap operator
given in Eq.~(\ref{eq:polargapoperator}) is that most of its matrix
elements $\Delta_{mj;m'j'}$ cancel. We can therefore calculate the
local DOS for reasonably large system sizes with moderate
computational effort.

In Fig.~\ref{fig:Dwave:xi=2.5s} we show results for the angular
average of the tunneling conductance $G({\bf r},\omega)$ for various
values of $r$ calculated for a system with radius $k_F R = 120$ and
a cutoff $m_{\text{max}} = 60$ for the angular momentum channels.
Directly at the vortex center there is a peak in the local DOS near
the Fermi level which is due to a continuum of unlocalized states
whose envelopes take there maximum near the vortex center but which
have tails leaking out in the nodal directions. These results are
consistent with those found in Refs.~\onlinecite{Wang95,Franz98}. As
we move away from the vortex center the peak in the local DOS
gradually vanishes and the tunneling conductance assumes the
familiar features of a $d$-wave bulk spectrum (at finite
temperature).

To calculate the transition matrix elements
$M_{\ell,\ell'}^{\alpha}$ which were defined in
Eq.~(15) of I
we will use the Hellmann-Feynman theorem
which in the present case of a $d$-wave gap operator reads
\begin{align}
  \label{eq:Hellmanndwave}
  M_{\ell,\ell'}^{+} & = (\epsilon_{\ell'} - \epsilon_{\ell}) U_{\ell;\ell'}^{+} - \frac{R_0}{2 m_e} \int_0^{2\pi} d\theta\, \partial_r \Psi_{\ell}^{\dagger} \sigma_3 \partial_{\bar z} \Psi_{\ell'} \Big|_{r = R_0} \nonumber \\
  & \quad {} + \frac{\Delta(R_0) R_0}{2 k_F^2} \int_0^{2\pi} d\theta\,
  (e^{3i\theta} + e^{-i\theta}) \partial_r u_{\ell}^{\ast} \partial_{\bar z} v_{\ell'} \Big|_{r = R_0} \nonumber \\
  & \quad {} + \frac{\Delta(R_0) R_0}{2 k_F^2} \int_0^{2\pi} d\theta\,
  (e^{-3i\theta} + e^{i\theta}) \partial_r v_{\ell}^{\ast} \partial_{\bar z} u_{\ell'} \Big|_{r = R_0} \;.
\end{align}
The first two terms are as in the $s$-wave case but since our
$d$-wave operator involves two derivatives, partial integration now
also leads to a boundary term involving the bulk gap. In the
$s$-wave case angular momentum was a good quantum number and for a
transition matrix element not to vanish we had the selection rule
that the angular momenta of the two wave functions had to differ by
one. As already emphasized above, in the $d$-wave case angular
momentum is not a good quantum number. Instead, each eigenstate
belongs to one of four sectors which only contains certain angular
momentum channels. As a result of this we find that the transition
matrix elements $M_{\ell,\ell'}^{+} = M_{\ell',\ell}^{-}$ are only
non-zero if $s'(\ell') = s(\ell) + 1\, (\text{mod}\, 4)$. In this
case we can again express all wave functions in terms of their
Fourier-Bessel components and obtain
\begin{widetext}
\begin{align}
  \label{eq:MKK2}
  M_{\ell,\ell'}^{+} = \,
  & \frac{1}{2}
  \sum_{m=4\nu+s} \sum_{j j'}
  c_{m j}^{(\ell)} \left[(\epsilon_{\ell'} - \epsilon_{\ell}) \mathcal{K}_{j j'}^{(m)} - \mathcal{L}_{j j'}^{(m)} \right] c_{m+1,j'}^{(\ell')}
  + \frac{1}{2}
  \sum_{m=4\nu+s-1} \sum_{j j'}
  d_{m j}^{(\ell)} \left[(\epsilon_{\ell'} - \epsilon_{\ell}) \mathcal{K}_{j j'}^{(m)} + \mathcal{L}_{j j'}^{(m)} \right] d_{m+1,j'}^{(\ell')}   \nonumber \\
 + & \frac{\Delta(R_0)}{2 k_F^2 R_0^3}
  \sum_{m=4\nu+s} \sum_{j j'} \left\{
  c_{m j}^{(\ell)} (-1)^{j-j'} \alpha_{mj} \alpha_{mj'} d_{m j'}^{(\ell')}
  + c_{m j}^{(\ell)} (-1)^{j-j'} \eta_{-1}(m)\, \eta_{-3}(m)\,  \alpha_{mj} \alpha_{m+4,j'} d_{m+4, j'}^{(\ell')} \right\}
\nonumber \\
 + & \frac{\Delta(R_0)}{2 k_F^2 R_0^3}
  \sum_{m=4\nu+s-1} \sum_{j j'} \left\{
  d_{m j}^{(\ell)} (-1)^{j-j'} \eta_{1}(m) \alpha_{mj} \alpha_{m-2,j'} c_{m-2, j'}^{(\ell')}
  + d_{m j}^{(\ell)} (-1)^{j-j'} \eta_{-1}(m) \,  \alpha_{mj} \alpha_{m+2,j'} c_{m+2, j'}^{(\ell')} \right\} \;,
\end{align}
\end{widetext}
where the matrix elements $\mathcal{K}_{j j'}^{(m)}$ and
$\mathcal{L}_{j j'}^{(m)}$ were already defined in
Eqs.~(40) and (41) of I
and $\eta_n(m) \equiv
1-2\delta_{n,m}$ leads to an extra minus sign if $m=n$.
\begin{figure}[tb]
  \includegraphics[width=8cm]{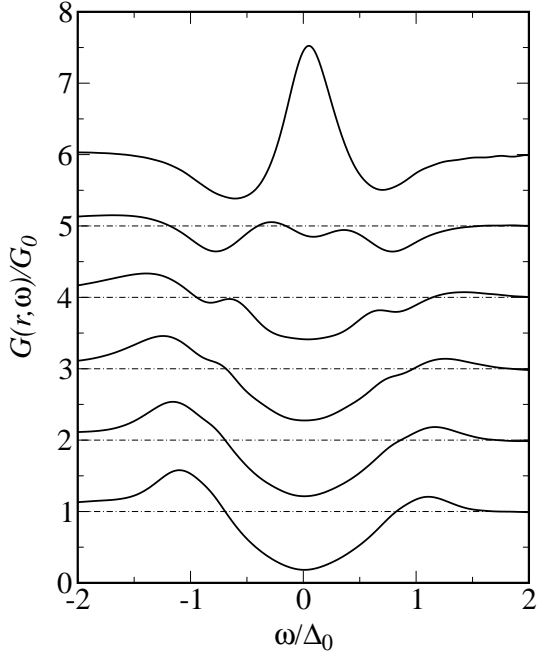}
  \caption{Angular average of the tunneling conductance $G=\partial I / \partial V$ for a superconductor with $d_{x^2-y^2}$ wave symmetry and a static vortex at the origin as a function of $\omega$. We have chosen $k_F \xi=2.5$ and $T=0.02\, E_F$. The upper curve is for $r = 0$ and the curves below are for $k_F r = 4,8,\dots,20$.}
\label{fig:Dwave:xi=2.5s}
\end{figure}
\begin{figure}[tb]
  \centering
  \includegraphics[width=8cm]{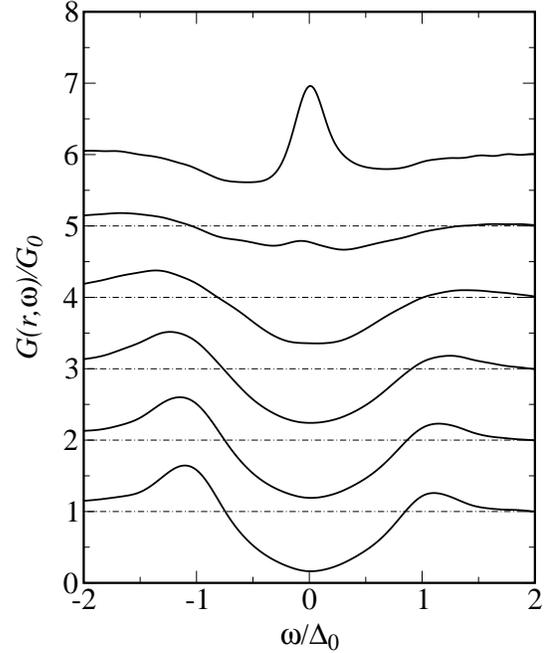}
  \caption{Angular average of the tunneling conductance $G=\partial I / \partial V$ for a superconductor with $d_{x^2-y^2}$ wave symmetry and a vortex with vortex mass $m_v = m_e$ as a function of $\omega$. We have chosen $k_F \xi=2.5$ and $T=0.02\, E_F$. The upper curve is for $r = 0$ and the curves below are for $k_F r = 4,8,\dots,20$.}
\label{fig:Dwave:xi=2.5M1}
\end{figure}

The tunneling conductance calculated for the parameters used above
but with a vortex mass equal to the mass of an electron is shown in
Fig.~\ref{fig:Dwave:xi=2.5M1}. As seen in experiments, the central
peak in the vortex center is suppressed and weight is shifted to
both sides away from the Fermi level. Although the mechanism for
redistributing spectral weight is the same as in the $s$-wave case
no satellite peaks are visible in the $d$-wave case. The
redistribution of spectral weight is further illustrated in
Fig.~\ref{fig:DwaveComp} where we compare our results for a static
vortex with those for a moving vortex.
\begin{figure}[tb]
  \centering
  \includegraphics[width=8cm]{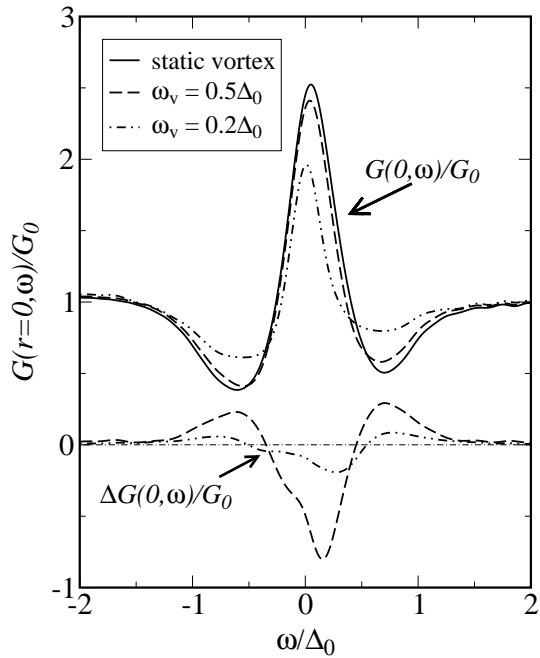}
  \caption{Comparison of the tunneling conductance $G=\partial I / \partial V$ for a superconductor with $d_{x^2-y^2}$ wave symmetry in the presence of a static or moving vortex.
The upper curve shows the tunneling conductance at the vortex center
for a static vortex (with $k_F \xi=2.5$ and $T=0.02\, E_F$). As we
allow for a finite vortex mass and vortex frequency the cental peak
decreases and weight is shifted away from the Fermi level. This is
shown in the two curves below for which we have chosen $m_v = m_e$
and $\omega_v = 0.2\,\Delta_0$ or $\omega_v = 0.5\,\Delta_0$
respectively. Finally, the two curves at the bottom of the figure
represent the difference between the results for the case of a
moving vortex and that of a static vortex.} \label{fig:DwaveComp}
\end{figure}
As can be seen in that figure, choosing a smaller vortex frequency
$\omega_v$ leads to a much stronger redistribution of spectral
weight. This is a direct consequence of
Eqs.~(16) and~(17) of I.
Also, although there are no sub-gap peaks visible
in the tunneling conductance, the curves for the difference between
the tunneling conductance for a moving and that for a static vortex
show peaks at an energy significantly smaller than the gap energy,
which according to Eq.~(16) of I
should be of the order of the vortex frequency.

Although our theory explains the redistribution of spectral weight
from the zero bias level towards higher energies, it cannot explain
the exact line shape as observed in experiment. First of all, the
system sizes studied are much smaller than in the $s$-wave case,
leading to a poorer spectral resolution. Angular momentum is not a
good quantum number and a large number of angular momentum channels
is needed to obtain convergence. This is shown in
Fig.~\ref{fig:angularmomentum} where we compare results for the
tunneling conductance at the vortex center, calculated with
different cutoffs $m_{\text max}$ implemented.
\begin{figure}[tb]
  \centering
  \includegraphics[width=8cm]{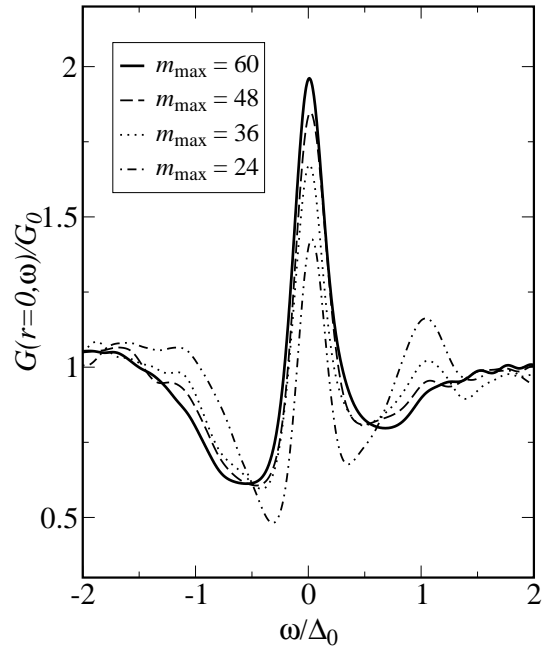}
  \caption{Comparison of results for the tunneling conductance
  $G=\partial I / \partial V$ at the vortex center of a superconductor
  with $d_{x^2-y^2}$ wave symmetry in the presence of a moving vortex.
  As in the previous figures, we have considered a system with radius
  $k_F R_0 = 120$, $k_F \xi=2.5$ and $T=0.02\, E_F$. The only difference
  is that we have used different cutoffs ($m_{\text{max}} = 24,\,36,\,48$, and $60$)
  for the angular momentum channels.}
\label{fig:angularmomentum}
\end{figure}
As can be seen in that figure, we have not even achieved convergence
for $m_{\text{max}} = 60$, the value used in all the previous
figures. However, it seems that we are not too far away from the
$m_{\text max} \to \infty$ limit. (It should also be noted that an
unphysical peak at $\approx +\Delta_0$ as seen for $m_{\text max} =
24$ or $36$ is basically absent for $m_{\text max} = 60$.)

The above scenario is quite different from the case of a static
vortex where a cutoff of $m_{\text max} = 24$ suffices to obtain
convergence. We therefore conclude that it must be the transition
matrix elements $M_{\ell,\ell'}^{\pm}$ which are very sensitive to
the boundary conditions imposed by the cutoff $m_{\text max}$.
Another restriction on the applicability of our theory is our
expansion of the gap operator $\hat \Delta ({\bf r} - {\bf
R}(\tau))$ to leading order in ${\bf R}(\tau)$. Since the quantum
zero-point motion of the vortex basically extends roughly over a
distance of $1/\sqrt{m_v \omega_v}$ from the vortex center this
approximation is only expected to be reasonable as long as $\xi \gg
1/\sqrt{m_v \omega_v}$. We are therefore not able to study systems
with very small coherence lengths.


\end{document}